\documentstyle[12pt]{article}
\begin{document}
\begin{titlepage}
\title{On the problem of chaos conservation
in quantum physics}

\author{
   V.P.Maslov and O.Yu. Shvedov  \\
{\small{\it Sub-faculty of Quantum Statistics and Field Theory,  }}\\
 {\small{\it Department of Physics, Moscow State University, }}\\
{\small{\it Vorobievy gory, Moscow 119899, Russia}} }

\end{titlepage}
\maketitle

\begin{center}
{\bf Abstract}
\end{center}

{\small
  We develop a new method of constructing an asymptotic series in powers of
$N^{-1/2}$ as $N\rightarrow\infty$ for the function of $N$ arguments which
is a solution to the Cauchy problem for the equation of a special type.
Many-particle Schr\"{o}dinger, Wigner and Liouville equations for a system of
a large number of particles are of this type, when the external potential is
of order $O(1)$, while the coefficient of the particle interaction potential
is $1/N$; the potentials can be arbitrary smooth bounded functions. We apply
this method to equations for $N$-particle states corresponding to the $N$-th
tensor power of an abstract Hamiltonian algebra of observables. In particular,
we show for the case of multiparticle Schr\"{o}dinger-like equations that the
property of $N$-particle wave function to be approximately equal at large $N$
to the product of one-particle wave functions does not conserve under time
evolution, while the same property for the correlation functions of the
finite order is known to conserve (such hypothesis being the quantum analog
of the chaos conservation hypothesis put forward by M.Kac in 1956 was proved
by the analysis of the BBGKY-like hierarchy of equations). In order to find a
leading asymptotics for the $N$-particle wave function, one should use not
only the solution to the well-known Hartree equation being derivable from the
BBGKY approach but also the solution to another (Riccati-type) equation
presented in this paper. We also consider another interesting case when one
adds to the $N$-particle system under consideration one more particle
interacting with the system with the coefficient of the interaction potential
of order $O(1)$. It happens that in this case one should investigate not a
single Hartree-like equation but a set of such equations, and the chaos will
not conserve even for the correlation functions.
}

\newpage
\section{Introduction}

The chaos conservation hypothesis is well-known in statistical physics.
This hypothesis put forward by M.Kac [1] in 1956 for the case of classical
systems has the following analog in the quantum case of the system of $N$
bose-particles moving in $\nu$-dimensional space. Consider the $k$-particle
correlation functions [2]
$$
{\cal R}^{t}_{k,N}(x_1,...,x_k;y_1,...,y_k)=
$$
\begin{equation}
\int dx_{k+1}...dx_N
\Psi^{t}_{N}
(x_1,...,x_k,x_{k+1},...,x_N)\Psi^{t*}_{N}(y_1,...,y_k,x_{k+1},...,x_N)
\end{equation}
corresponding to $N$-particle wave functions $\Psi^{t}_{N}(x_1,...,x_N)$
which specify states of the system and satisfy the
$N$-particle Schr\"{o}dinger
equation ($x_1,...,x_N \in {\bf R}^{\nu}$ are particle coordinates,
$t \in {\bf R}$ is the time variable).The quantum analog of the chaos
hypothesis is the following. Suppose that at the initial instant of time $t=0$
the correlator (1) factorizes as $N\rightarrow\infty,k=const$ as follows:
\begin{equation}
{\cal R}^{t}_{k,N}(x_1,...,x_k;y_1,...,y_k)
 \rightarrow \varphi^t(x_1)...
\varphi^t(x_k)\varphi^{t*}(y_1)...\varphi^{t*}(y_k),
\end{equation}
where $\varphi^t$ is one-particle wave function
such that $\int dx |\varphi^t(x)|^2=1$. Then the property (2) holds
for arbitrary time $t$ as well.

The discussed hypothesis can be justified for the case of the external
potential of order $O(1)$ and the particle interaction potential of order
$O(1/N)$. The mathematical proof has been obtained in [3].The method
of justification of the property (2) was based on the ideas of [2]:
the $N$-particle equation for the density matrix
$
\Psi^{t}_{N}(x_1,...,x_N)\Psi^{t*}_{N}(y_1,...,y_N)
$
was integrated over last $N-k$ variables and the chain of equations for
${\cal R}^t_{k,N}$ was obtained in a way analogous to the method of
derivation of the BBGKY hierarchy found almost simultaneously by
Bogoliubov, Born, Green, Kirkwood and Yvon for the classical case. It was
also found in [2,3] that the function $\varphi^t$ obeys the Hartree equation
being of widely use in physics for studying quantum systems with a large
number of particles.

The property (2) has the following physical meaning in terms of mean values
of the observables being operators acting in Hilbert space
$L^2({\bf R}^{\nu N}).$ Consider observables $A_N$ with kernels of the
special form
$$
A_N(x_1,..,x_N;y_1,...,y_N)=\sum_{p=1}^{P_0} \frac{1}{N^{p}p!}
\sum_{1\le i_1 \ne ... \ne i_p \le N}
$$
\begin{equation}
A^{(p)}(x_{i_1},...,x_{i_p};y_{i_1},...,y_{i_p}) \prod_{i\ne i_l}
\delta(x_i-y_i),
\end{equation}
where $P_0 \in {\bf N}$,$A^{(p)}$ are kernels of operators acting in
$L^2({\bf R}^{\nu p})$.The property (2) implies that the mean values of
the observables $A_N$  in the state $\Psi^t_N$ have the following limit as
$N \rightarrow \infty$:
$$
(\Psi^t_N,A_N\Psi^t_N)
\equiv \int dx_1...dx_N \Psi^{t*}_N(x_1,...,x_N)
(A_N\Psi_N^t)(x_1,...,x_N)
\rightarrow
$$
$$
\rightarrow
\sum_{p=1}^{P_0}
\frac{1}{p!} \int \varphi^t(x_1)...
\varphi^t(x_k)\varphi^{t*}(y_1)...\varphi^{t*}(y_k) \times
$$
$$
 \times A^{(p)}(y_1,...,y_p;x_1,...,x_p)dx_1...dx_p dy_1...dy_p.
$$
In this paper we consider a new formulation of a problem. Namely, we discuss
if the property (2) is valid when $k$ also tends to infinity, for
example, $k=N$. We also study whether the chaos property allows us to find
limits as $N\rightarrow \infty$ of mean values of  observables of a more
general form than (3).

To solve these problems, we construct an asymptotic formula as $N\rightarrow
\infty$ not for the  correlators (1)
 but for the full $N$-particle wave function obeying the initial
condition of a product of one-particle wave functions
  (remind that a number of arguments of the wave function also tends
 to infinity as the parameter of the asymptotic expansion tends to zero).
 We will see that the asymptotics will not factorize into a product of
 one-particle wave functions. Therefore, the chaos hypothesis fails if
 $k\rightarrow\infty$ in (1). Thus, when one makes an attempt to find
 mean values of general observables uniformly bounded with respect to $N$,
 one can't use the property of factorizing of the wave function, contrary
 to the case of the observables of the type (3).

Consider the multiparticle Schr\"{o}dinger equation
$$
i\hbar \frac{\partial}{\partial t} \Psi^t_N(x_1,...,x_N)=
$$
\begin{equation}
\left[\sum_{i=1}^{N}\left(-\frac{\hbar^2}{2m} \Delta_i + U(x_i)\right)
+\frac{1}{N} \sum_{1 \le i<j \le N} V(x_i,x_j) \right] \Psi^t_N(x_1,...,x_N),
\end{equation}
where $\hbar$ is the Planck constant, $m$ is the particle mass, $\Delta_i=
\partial^2/\partial x_i^2$ is the Laplace operator, $U$ is the external
potential, $\frac{1}{N}V$ is the particle interaction potential being of order
$1/N$.
The asymptotic formula for the $N$-particle wave function is then expressed
not only through the solution to the well-known Hartree equation
\begin{equation}
i\hbar \frac{\partial}{\partial t} \varphi^t(x)=
\left[-\frac{\hbar^2}{2m} \Delta + W^t(x)\right] \varphi^t(x),
\end{equation}
where $W^t$ is a self-consistent potential
\begin{equation}
W^t(x)=U(x)+\int V(x,y) |\varphi^t(y)|^2 dy.
\end{equation}
One should also use the following system:
$$
i\hbar \frac{\partial}{\partial t} u^t(x)=
\left[-\frac{\hbar^2}{2m} \Delta + W^t(x)\right] u^t(x)
$$
$$
+\varphi^t(x)\int dy V(x,y)(\varphi^{t*}(y)u^t(y)+v^t(y)\varphi^t(y)),
$$
\begin{equation}
\end{equation}
$$
-i\hbar \frac{\partial}{\partial t} v^t(x)=
\left[-\frac{\hbar^2}{2m} \Delta + W^t(x)\right] v^t(x)
$$
$$
+\varphi^{t*}(x)\int dy V(x,y)(\varphi^{t*}(y)u^t(y)+v^t(y)\varphi^t(y))
$$
The system (7) can be formally obtained by the following procedure. One
can write the system consisting of the Hartree equation (5)
 and the equation conjugated to it, consider the variation system for it and
 substitute the variations of $\varphi$ and $\varphi^{*}$ by $u$ and $v$
 that should not be conjugated. This variation system with independent
 variations of $\varphi$ and $\varphi^{*}$ coincides with (7).
The asymptotic formula for the $N$-particle wave function is expressed
through the solution to eq.(5) and through the operator transforming
the initial condition for the Cauchy problem for system (7) into the
solution to this Cauchy problem.

We can consider equations of a more general form than the multiparticle
\newline
Schr\"{o}dinger equation (4). Namely, we can study equations for the functions
$\Psi_N \in L^2({\cal X})$, where $\cal X$ is an arbitrary measure space.
Such equations are of the form
$$
i\frac{d}{dt} \Psi_N^t = N A_N \Psi_N^t,
$$
where operator $A_N$ has a kernel of the type (3). Notice that the
Schr\"{o}dinger equation is a partial case of this equation.
 Asymptotic solutions to it
that obey more general initial conditions than a product
of one-particle wave functions are constructed in section 5. These asymptotic
formulas implying  the results on the chaos non-conservation
for eq.(4) are proved
in section 6. In section 7 we evaluate the corrections to the asymptotic
formula.

We can consider not only Schr\"{o}dinger-like equations but also sets of such
equations. The generalization of the method which is applicable to such case
is to be developed in sections 8,9. The technique of these sections can be also
used when one considers the $N$-particle systems interacting with an additional
particle, so that the evolution equation has the form:
$$
i\hbar \frac{\partial}{\partial t} \Psi^t_N(y,x_1,...,x_N)=
[ N\left(-\frac{\hbar^2}{2M}\Delta_y+{\cal U}(y)\right)
+\sum_{i=1}^N {\cal V}(x_i,y)
$$
\begin{equation}
+\sum_{i=1}^{N}\left(-\frac{\hbar^2}{2m} \Delta_i + U(x_i)\right)
+\frac{1}{N} \sum_{1 \le i<j \le N} V(x_i,x_j)] \Psi^t_N(y,x_1,...,x_N),
\end{equation}
The first term of the right-hand side of this equation corresponds to the
 motion  of the additional particle of the mass $M/N$ in the external
potential ${\cal U}(y)$ ($y$ is the coordinate of the additional particle).
The second term corresponds to the interaction of the additional particle with
the $N$-particle system. If the first and the second terms were of orders
$O(1)$ and $O(1/N)$ correspondingly, one could apply the technique
 to be considered in sections 5-7 without modification. But the coefficients
are $N$ and 1, so another approach is needed. It happens that the Hartree
equation (5) should be modified, there will be no longer a single
Hartree-like equation, there will be a set of such equations, and the analog
of the chaos hypothesis will then fail even for the correlation functions.

We can consider the problem of chaos conservation not only for $N$-particle
wave functions obeying the multiparticle Schr\"{o}dinger equation but also
for other cases. Namely, one can investigate $N$-particle density metrices
obeying the $N$-particle
\newline
Wigner equation
 [13] or $N$-particle density functions (probability distributions)
which obey the multiparticle Liouville equation [11].
One can find [11,13] asymptotic formulas for these cases. These asymptotics
are also products of one particle densities at the initial time moment,
while there is no such factorization at time moment $t$. In section 10
we will consider the equations for $N$-particle states corresponding
to the $N$-th tensor power of an abstract Hamiltonian algebra of observables.
We generalize a notion of a half-density (discussed in [11,13] for
different cases) by introducing the notions of a half-density representation
of an abstract Hamiltonian algebra and of an abstract half-density.
There are following examples of the latter notion:

(a) the square root of the $N$-particle probability distribution [11];

(b) the square root of the $N$-particle density matrix [13];

(c) the $N$-particle wave function.

We show, that when our asymptotic method is applied to the $N$-particle
half-density equation, the average values of general bounded observables
is unambiguously defined by our approximation for the half-density.
The results to be obtained in section 10 imply, in particular, the
asymptotic formulas found for the cases of Schr\"{o}dinger,
Liouville, Wigner equations.

\section{Violation of the chaos hypothesis for the N-particle wave function}

{\bf 1}.In the previous section we have seen that if the property (2) is
satisfied, one can replace the wave function $\Psi^t_N$ by the product
of one-particle wave functions $\varphi^t(x_1)...\varphi^t(x_N)$ in order
to find a limit as $N\rightarrow\infty$ of the mean values of the observables
of the special form (3). Consider the problem if such replacement
is valid for finding mean values of general observables $A_N$
uniformly bounded with respect to $N$, $||A_N||<C$. What is necessary and
sufficient condition for this replacement? The following lemma tells us
that such condition is
\begin{equation}
\int dx_1...dx_N |\Psi^t_N(x_1,...,x_N)- c_N^t \varphi^t(x_1)...
\varphi^t(x_N)|^2 \rightarrow_{N\rightarrow\infty} 0
\end{equation}
for some number $c_N^t \in {\bf C}$,
$|c_N^t|\rightarrow_{N\rightarrow\infty} 1.$

{\bf Lemma 1.}
{\it Let $\Phi_N$ and $\Psi_N$ be such elements of $L^2({\bf R}^{\nu N})$
     that $(\Phi_N,\Phi_N)=(\Psi_N,\Psi_N)=1.$

     1.Let $A_N$ be operators acting in $L^2({\bf R}^{\nu N})$ and
       uniformly bounded with respect to $N$,$||A_N||<C$. Let for some
       set of numbers
       $c_N \in {\bf C},|c_N| \rightarrow_{N\rightarrow\infty} 1$
\begin{equation}
||\Psi_N - c_N\Phi_N|| \rightarrow_{N\rightarrow\infty} 0
\end{equation}
       Then the property
       $
(\Phi_N,A_N \Phi_N) \rightarrow_{N\rightarrow\infty} A
       $
       implies that
       $
(\Psi_N,A_N \Psi_N) \rightarrow_{N\rightarrow\infty} A
       $

      2.Let
\begin{equation}
(\Psi_N,A_N \Psi_N) -
(\Phi_N,A_N \Phi_N) \rightarrow_{N\rightarrow\infty} 0
\end{equation}
        for arbitrary set of operators $A_N$ uniformly bounded with respect
        to $N$. Then the property (10) is satisfied for some number
       $c_N \in {\bf C},|c_N| \rightarrow_{N\rightarrow\infty} 1$
 }

{\bf Proof.}

1. Denote $X_N=\Psi_N-c_N\Phi_N$. We have
$$
(\Psi_N,A_N \Psi_N) - (\Phi_N,A_N \Phi_N)
$$
$$
=c_N(X_N,A_N\Phi_N)+c_N^{*}(\Phi_N,A_N X_N) + (X_N,A_N X_N)
\rightarrow_{N\rightarrow\infty} 0
$$
because $||A_N||<C$ and $||X_N|| \rightarrow_{N\rightarrow\infty} 0$.

2. Consider the sequence $X_N$ of the form
$$
X_N=\Psi_N-\Phi_N(\Phi_N,\Psi_N).
$$
Notice that $(\Phi_N,X_N)=0$ and choose the following set of operators $A_N$:
$$
A_N w = X_N(X_N,w), w \in L^2({\bf R}^{\nu N})
$$
that are uniformly bounded with respect to $N$: namely,
$$
||A_N||=||X_N||^2 \le (||\Psi_N||+||\Phi_N||)^2=4.
$$
It follows from eq.(11) that
$$
(\Psi_N,A_N \Psi_N) - (\Phi_N,A_N \Phi_N)
=|(X_N,\Psi_N)|^2-
|(X_N,\Phi_N)|^2=
$$
$$
= |(X_N,X_N)|^2
\rightarrow_{N\rightarrow\infty} 0,
$$
so that the property (10) is satisfied for $c_N=(\Phi_N,\Psi_N).$
As $1-|c_N|^2=
(\Psi_N-c_N\Phi_N,\Psi_N-c_N\Phi_N)\rightarrow_{N\rightarrow\infty} 0$,
one has $|c_N|\rightarrow_{N\rightarrow\infty} 1$.
Lemma 1 is proved.

{\bf 2}.Let $U,V$ be smooth functions $U:{\bf R}^{\nu} \rightarrow {\bf R},
V:{\bf R}^{\nu}\times {\bf R}^{\nu} \rightarrow {\bf R}$ bounded with all
their derivatives, $V(x,y)=V(y,x).$ Denote by
$W_2^{\infty}({\bf R}^{\nu})$ the following set:
$$
W_2^{\infty}({\bf R}^{\nu})=\{f:{\bf R}^{\nu} \rightarrow {\bf C} |
 f \in  W_2^k({\bf R}^{\nu}), k=\overline{1,\infty}  \}
$$
Consider the initial condition $\varphi^0:{\bf R}^{\nu}
\rightarrow {\bf C}$ for the
Hartree equation (5) such that
\begin{equation}
\varphi^0 \in W_2^{\infty}({\bf R}^{\nu}), \int |\varphi^0(x)|^2 dx=1.
\end{equation}
The following lemma is proved in [4,5].

{\bf Lemma 2.}
{\it There exists a unique solution
$\varphi^t \in W_2^{\infty}({\bf R}^{\nu})$
to the Cauchy problem for eq.(5).}

As the Schr\"{o}dinger equation is the partial case of the Hartree equation,
lemma 2 implies the following corollary.

{\bf Corollary.} {\it There exists a unique solution
$\Psi^t_N \in W_2^{\infty}({\bf R}^{\nu N})$
to eq.(4) which satisfies the initial condition}
\begin{equation}
\Psi_N^0(x_1,...,x_N)=\varphi^0(x_1)...\varphi^0(x_N).
\end{equation}

It occurs that the property (9) is not valid.

{\bf Theorem 1}.
{\it
  Let $V(x,y) \ne 0$ for any $x,y \in {\bf R}^{\nu}.$
  Then there is no such interval $[t_1,t_2]$ that the property (9) is
  satisfied for $t \in [t_1,t_2]$ for some number $c_N^t \in {\bf C}.$
 }

This theorem is a corollary of the more general statement to be proved in
section 6.

{\bf 3.} Let us consider a heuristic method to derive the result of
 theorem 1. Suppose that for some function $c_N^t$ the initial condition (13)
 evolve into the wave function
\begin{equation}
c_N^t\varphi^t(x_1)...\varphi^t(x_N)+z^t_N(x_1,...,x_N),
\end{equation}
where $||z_N||_{L^2} \rightarrow_{N\rightarrow\infty} 0$. One can then
expect that the property of chaos conservation for the full $N$-particle
wave function is also valid when the initial condition for the Hartree
equation (5) is shifted by the quantity of order $1/\sqrt{N}$:
$$
\varphi^0 \rightarrow \varphi^0 +\delta\varphi^0/\sqrt{N}.
$$
In order to retain the property of the norm of the wave function to be
of order $O(1)$, let us choose the variation $\delta\varphi^0$ to be
orthogonal to $\varphi^0$. As the Hartree equation contains not only
$\varphi^t$ but also $\varphi^{t*}$, the function $\varphi^t$ transforms
as follows:
\begin{equation}
\varphi^t \rightarrow \varphi^t +\frac{1}{\sqrt{N}}
(A^t \delta\varphi^0 +
B^t (\delta\varphi^0)^{*}) + O(1/N),
\end{equation}
where $A^t$ and $B^t$ are some linear operators acting in
$L^2({\bf R}^{\nu}).$
Applying the transformation (15) to formula
(14) and multiplying $\delta\varphi^0$
by $e^{ia}$, one obtains that the following $N$-particle wave function
$$
c_N^t
(\varphi^t(x_1) +\frac{1}{\sqrt{N}}
(A^t \delta\varphi^0(x_1)e^{ia} +
B^t (\delta\varphi^0)^{*}(x_1)e^{-ia}) + ...)
$$
\begin{equation}
\times ...
(\varphi^t(x_N) +\frac{1}{\sqrt{N}}
(A^t \delta\varphi^0(x_N)e^{ia} +
B^t (\delta\varphi^0)^{*}(x_N)e^{-ia}) + ...)
+z^t_N(x_1,...,x_N)
\end{equation}
is also an asymptotic solution to eq.(4). At the initial time moment formula
(16) does not contain negative powers of $e^{ia}$,
while at time moment $t$ such powers arises. For example,the contribution of
the power $e^{-ia}$ is zero at initial time moment and is equal to
$$
a_N^t\varphi^t(x_1)...\varphi^t(x_N)
$$
\begin{equation}
+c_N^{t'} \frac{1}{\sqrt{N}} \sum_{i=1}^{N}
\varphi^t(x_1)
..\varphi^t(x_{i-1})
B^t (\delta\varphi^0)^{*}(x_i)
\varphi^t(x_{i+1})
..\varphi^t(x_N)
\end{equation}
at time moment $t$ for some numbers $a_N^t,c^{t'}_N \in {\bf C}$
(the term with $a_N^t$ arises from the coefficient $c_N^t$ that may be
$a$-dependent). Because of the linearity of eq.(4), the $N$-particle wave
function (17) is also expected to be asymptotic solution to eq.(4).
Notice that the function
$B^t (\delta\varphi^0)^{*}$
can be decomposed into two parts: one of them being proportional to
$\varphi^t$ and another being orthogonal to $\varphi^t$:
$$
B^t (\delta\varphi^0)^{*}=
b^t \varphi^t +
B^{t'} (\delta\varphi^0)^{*}
$$
The contribution of the term $b^t\varphi^t$ to eq.(17) can be involved to
the first term of eq.(17),
while another term can't be treated in this way. As
the norm of its contribution
is of order $O(1)$, we are faced with the difficulty: the $N$-particle
wave function being equal to zero at the initial time moment
evolves into non-zero wave function.
The only possible way to resolve the difficulty
 is to adopt the violation of the
chaos hypothesis (9) when
$B^{t'} \ne 0$. The theorem 1 is heuristically justified.

One can also expect that investigation of the operators $A^t,B^t$ being
obtainable from the variation system (7) can lead us to the correct
asymptotic formula for the wave function. This is to be done in the
following sections.

\section{Multiparticle canonical operator and asymptotic formula
for the $N$-particle wave function as $N \rightarrow\infty$}

{\bf 1.} In the previous section we have seen that the wave function (17)
may play an important role in constructing the $N$-particle wave function
asymptotics as $N\rightarrow\infty.$ We can also notice that such wave
function
satisfies the chaos property (2) and does not satisfy the propery (9).
 Therefore, the wave function (17) gives us an example of the
state which can be replaced by the product
$\varphi^t(x_1)...\varphi^t(x_N)$
in order to find limits of mean values of observables of the special
form (3) but not of the general form.

Let us give a generalization of the example (17). Introduce a notion of a
multiparticle canonical operator being a partial case of the canonical
operator corresponding to Lagrangian manifold with complex germ in
Fock space [6,7].

Let us introduce the following notations. By $\cal F$ we denote the
space of sets $(g_0,g_1(x_1),g_2(x_1,x_2),...)$ of functions
$g_n:{\bf R}^{\nu n} \rightarrow {\bf C}$ which are symmetric with respect to
$x_i \in {\bf R}^{\nu}$, belong to the space
$L^2({\bf R}^{\nu n})$
 and satisfy the condition
\begin{equation}
\sum_{n=0}^{\infty} \int dx_1...dx_n |g_n(x_1,...,x_n)|^2 < \infty.
\end{equation}
By $g_n$ we denote the $n$-th component of $g \in {\cal F}$. Define an inner
product in $\cal F$ as
$$
(g^{(1)},g^{(2)})=
\sum_{n=0}^{\infty} \int dx_1...dx_n
g_n^{(1)*}(x_1,...,x_n)
g_n^{(2)}(x_1,...,x_n).
$$
By ${\cal F}_{\varphi}$,
$\varphi \in L^2({\bf R}^{\nu})$,
 we denote the subspace of $\cal F$ which
consists of all the elements $g \in {\cal F}$ such that
\begin{equation}
 \int dx_1...dx_n \varphi^{*}(x_1) g_n(x_1,...,x_n)=0, n=\overline{1,\infty}.
\end{equation}
Consider the following element of
$ L^2({\bf R}^{\nu N})$:
\begin{equation}
(K^N_{\varphi}g)(x_1,...,x_N)=\sum_{p=0}^{N}
\frac{\sqrt{p!}}{\sqrt{N^p}} \sum_{1\le i_1<...< i_p \le N}
g_p(x_{i_1},...,x_{i_p})\prod_{i \ne i_l} \varphi(x_i) ,
  \end{equation}
where $ g \in {\cal F}_{\varphi}.$

{\bf Remark.} The wave function (20) satisfies the chaos property (2).

{\bf Definition 1.}{\it An operator
$K^N_{\varphi}:{\cal F}_{\varphi} \rightarrow
 L^2({\bf R}^{\nu N})$ of the form (20) will be referred to as
 a multiparticle canonical operator.}

The following lemma shows that the norm of the function (20) is of order
$O(1)$, although there are $N!/(p!(N-p)!)$ terms in the sum over
$i_1,...,i_p$, while the coefficient of this sum is of order $N^{-p/2}$,
not of order $O(N^{-p})$.

{\bf Lemma 3.}
{\it The following relation
\begin{equation}
||K^N_{\varphi} g||^2 = \sum_{p=0}^N \frac{N!}{N^p(N-p)!}
\int dx_1...dx_p |g_p(x_1,...,x_p)|^2
\end{equation}
is satisfied.
}

{\bf Proof.} One has
$$
(K^N_{\varphi}g,K^N_{\varphi} g)
=\sum_{p=0}^{N} \frac{\sqrt{p!q!}}{\sqrt{N^p}}
\sum_{1\le i_1<...< i_q \le N}
\sum_{1\le j_1<...< j_p \le N}
$$
\begin{equation}
\int dx_1...dx_N g_q^{*}(x_{i_1},...,x_{i_q})
g_p(x_{j_1},...,x_{j_p})
\prod_{i \ne i_l}\varphi^{*}(x_i)
\prod_{j \ne j_m}\varphi(x_j).
\end{equation}
It follows from eq.(19) that for all non-vanishing terms in the sum (21)
$p=q$,while sets $i_1,...,i_p$ and $j_1,...,j_q$ coincide. As the number
of sets $i_1,...,i_p,1 \le i_1<...<i_p \le N$ is equal to $N!/(p!(N-p)!)$,
eq.(21) is satisfied. Lemma 3 is proved.

{\bf Remark.} We can notice that the main contribution to the inner
product
\newline
$(K^N_{\varphi}g,K^N_{\varphi}g)$ is given by the terms of the
sum over $p$ in eq.(20) which numbers are of order $O(1)$, not of
order $O(N).$

{\bf Corollary 1.}
$ ||K^N_{\varphi} g||^2 \le (g,g).$

{\bf Corollary 2.}
$ ||K^N_{\varphi} g||^2 \rightarrow_{N\rightarrow\infty} (g,g).$

{\bf Corollary 3.}
Let $f,g$ be such non-zero elements of ${\cal F}_{\varphi}$ that for
some $c_N \in {\bf C}$
$$
||K^N_{\varphi} f - c_N K^N_{\varphi} g||
\rightarrow_{N\rightarrow\infty} 0.
$$
Then for some $c \in {\bf C}$ $f=cg$.

{\bf 2.} It occurs that the approximate solution to eq.(4) which satisfies
the initial condition (13) should be expressed not only through the solution
$\varphi^t$ to the Hartree equation but also through the solution to another
equation. This is a Riccati-type equation:
$$
i\hbar \frac{\partial}{\partial t} R^t(x,y) =
V(x,y)\varphi^t(x)\varphi^t(y) +
\left(-\frac{\hbar^2}{2m}\Delta_x-\frac{\hbar^2}{2m}\Delta_y
+W^t(x)+W^t(y) \right)R^t(x,y)
$$
$$
+ \int dy^{'} \varphi^{t}(y)
V(y,y^{'})\varphi^{t*}(y^{'})R^t(x,y^{'}) +\int dx^{'} \varphi^{t}(x)
V(x,x^{'})\varphi^{t*}(x^{'})R^t(y,x^{'})
$$
\begin{equation}
+\int dx^{'} dy^{'}
R^t(x,x^{'}) R^t(y,y^{'})V(x^{'},y^{'})
\varphi^{t*}(x^{'})\varphi^{t*}(y^{'}),
 \end{equation}
where $W^t$ has
the form (6).

By $W_{\varphi}$ we denote the space of complex functions $R(x,y):
{\bf R}^{\nu} \times {\bf R}^{\nu} \rightarrow {\bf C}$ such that

(i) $R \in W_2^{\infty}({\bf R}^{2\nu}), R(x,y)=R(y,x);$

(ii)
\begin{equation}
\int R(x,y)\varphi^{*}(y) dy = -\varphi(x);
\end{equation}

(iii) the operator $M$ in $L^2({\bf R}^{\nu})$ with the kernel
\begin{equation}
M(x,y)=R(x,y) + \varphi(x)\varphi(y)
\end{equation}
satisfy the property $||M||<1.$

{\bf Lemma 4.} {\it Let $R^0 \in W_{\varphi^0}.$  Then there exists a
solution
$R^t \in W_{\varphi^t}$
to the Cauchy problem for eq.(23) with the initial condition $R^0$.
}

In order to prove this lemma, we will express the function $R^t$ through
the evolution operator transforming the initial condition for variation
system (7) into the solution to this system.

Let us first construct this operator and study its properties.
Denote by $L,Y^t$ and $Z^t$ the following operators in
$L^2({\bf R}^{\nu})$:
$$
L=-\frac{\hbar}{2m}\Delta,
(Z^tw)(x)=\hbar^{-1} \int dy V(x,y) \varphi^t(x)\varphi^t(y)w(y),
$$
\begin{equation}
\end{equation}
$$
(Y^tw)(x)=\hbar^{-1}[W^t(x)w(x)+\int dy V(x,y) \varphi^t(x)\varphi^{t*}(y)
w(y)].
$$
The solution to the Cauchy problem for the system (7) is then as follows:
\begin{equation}
\left(\begin{array}{c} u^t \\ v^t \end{array} \right)=
\left(\begin{array}{cc}  A^t & B^t \\ B^{t*} & A^{t*} \end{array}\right)
\left(\begin{array}{c} u^0 \\ v^0 \end{array} \right),
\end{equation}
where $A^t,B^t$ are operators in $L^2({\bf R}^{\nu})$ that satisfy the
following equation
\begin{equation}
i\frac{d}{dt}
\left(\begin{array}{cc}  A^t & B^t \\ B^{t*} & A^{t*} \end{array}\right)=
\left[
\left(\begin{array}{cc}  L & 0 \\ 0 & -L \end{array}\right)+
\left(\begin{array}{cc}  Y^t & Z^t \\ -Z^{t*} & -Y^{t*} \end{array}\right)
\right]
\left(\begin{array}{cc}  A^t & B^t \\ B^{t*} & A^{t*} \end{array}\right)
\end{equation}
and the initial condition
$$
A^0=E,B^0=0.
$$
We are going to show that the solution to eq.(28) is the following:
\begin{equation}
\left(\begin{array}{cc}  A^t & B^t \\ B^{t*} & A^{t*} \end{array}\right)=
\left(\begin{array}{cc}  e^{-iLt} & 0 \\ 0 & e^{iLt} \end{array}\right)
 \sum_{n=0}^{\infty}
\left(\begin{array}{cc}  A^t_n & B^t_n \\ B^{t*}_n & A^{t*}_n
\end{array}\right),
\end{equation}
where operators $A^t_n,B^t_n$ are determined from the following recursive
relations:
$$
\left(\begin{array}{cc}  A^t_0 & B^t_0 \\ B^{t*}_0 & A^{t*}_0
 \end{array}\right)=
\left(\begin{array}{cc}  E & 0 \\ 0 & E \end{array}\right)
$$
\begin{equation}
\left(\begin{array}{cc}  A^t_n & B^t_n \\ B^{t*}_n & A^{t*}_n
\end{array}\right)
=-i\int_0^t d\tau
\left(\begin{array}{cc}  Y^{\tau}_i & Z^{\tau}_i \\
-Z^{\tau *}_i & -Y^{\tau *}_i \end{array}\right)
\left(\begin{array}{cc}  A^{\tau}_{n-1} & B^{\tau}_{n-1} \\
B^{\tau *}_{n-1} & A^{\tau *}_{n-1}\end{array}\right),
\end{equation}
$$
Y^t_i=e^{iLt}Y^t e^{-iLt}, Z^t_i=e^{iLt}Z^t e^{-iLt}.
$$
Notice that the series (29) is well-defined, since the functions $U(x)$
and $V(x,y)$, as well as the operators $Y^t_i,Z^t_i$ are bounded.
The estimation
$$
||A_n^t|| \le \frac{c^nt^n}{n!}, ||B_n^t|| \le \frac{c^n t^n}{n!}
$$
for some constant $c$ can be justified by induction. Thus, the series (29)
converges.

 The following property is satisfied for operators (29).

 {\bf Lemma 5}.{\it
 The kernels of operators $B^t$ and $A^tR^0$ belong to the space
 $W_2^{\infty}({\bf R}^{2\nu})$.}

{\bf Proof.} Introduce notations:
$$
||S||=\sup_{||\xi||=1} ||S\xi||, ||S||_2 =\sqrt{ Tr S^+ S},
$$
$$
||S||^{(m)}=||(-\Delta+1)^m S (-\Delta+1)^{-m}||,
$$
$$
||S||^{(m)}_2=||(-\Delta+1)^m S (-\Delta+1)^{m}||_2
$$
for each operator $S$ in $L^2({\bf R}^{\nu}).$ Let us show by induction
that
\begin{equation}
||A_n^t||^{(m)} \le \frac{c_m^nt^n}{n!},
||B_n^t||^{(m)}_2 \le \frac{c_m^n t^n}{n!}
\end{equation}
for some constants $c_m$. Inequalities (31) are correct if $n=0.$ Suppose
them to be correct as $n<k$ and check eq.(31) as $n=k$. It follows from
 eq.(30) that
 $$
||A^t_k||^{(m)} \le \int_0^t d\tau [
||Y_i^{\tau}||^{(m)}||A^{\tau}_{k-1}||^{(m)} +
||Z_i^{\tau}||^{(m)}_2||B^{\tau}_{k-1}||^{(m)}_2 ],
 $$
 $$
||B^t_k||^{(m)}_2 \le \int_0^t d\tau [
||Y_i^{\tau}||^{(m)}||B^{\tau}_{k-1}||^{(m)}_2 +
||Z_i^{\tau}||^{(m)}_2||A^{\tau}_{k-1}||^{(m)} ].
 $$
 For $c_m>\max [ ||Y^{\tau}_i||^{(m)},
 ||Z^{\tau}_i||^{(m)}_2]$, eq.(31) is justified.

It follows from eq.(31) that
$$
\sum_{n=0}^{\infty} ||A^t_n||^{(m)} < \infty,
\sum_{n=0}^{\infty} ||B^t_n||^{(m)}_2 < \infty.
$$
Therefore,
\begin{equation}
 ||B^t||^{(m)}_2 < \infty,
 ||A^t||^{(m)} < \infty,
\end{equation}
$$
 ||A^tR^0||^{(m)}_2 \le
 ||A^t||^{(m)}
 ||R^0||^{(m)}_2 < \infty,
$$
Lemma 5 is proved.

 Let us show now that the operator (29) is really the solution to eq.(28).

{\bf Lemma 6.} {\it The following relations are satisfied;}
$$
||\left( i\frac{A^{t+\delta t}-A^t}{\delta t} -LA^t
 - (Y^tA^t+Z^tB^{t*})\right) R^0||_2
\rightarrow_{\delta t \rightarrow 0} 0
$$
$$
||i\frac{B^{t+\delta t}-B^t}{\delta t} -LB^t - (Y^tB^t+Z^tA^{t*})||_2
\rightarrow_{\delta t \rightarrow 0} 0
$$

{\bf Proof.} It follows from eqs.(29),(30) that
$$
\frac{A^{t+\delta t}-A^t}{\delta t} +iLA^t +i
(Y^tA^t+Z^tB^{t*})=
\frac{e^{-iL\delta t}-1 + iL\delta t}{\delta t} A^{t}
$$
$$
+(1-e^{-iL\delta t})i(Y^tA^t+Z^tB^{t*})
$$
$$
-e^{-iL(t+\delta t)}
\frac{i}{\delta t} \int_0^{\delta t} d\tau \sum_{n=0}^{\infty}
(
Y_i^{t+\tau}A_n^{t+\tau}+Z_i^{t+\tau}B_n^{t+\tau *}
-Y_i^{t}A_n^{t}-Z_i^{t}B_n^{t *}
).
$$
Making use of eq.(32) and of the relation
$$
e^{-iL\tau}-1=\int_0^{\tau} d\alpha e^{-iL\alpha} (-iL),
$$
we find that the first statement of lemma 6 is satisfied. The proof of the
second statement is analogous. Lemma 6 is proved.

Let us construct the solution to eq.(23).

{\bf Lemma 7.}
{\it

1.The operator
 $A^{t*}+B^{t*}R^0$
 is boundedly invertable.

2. The kernel of the operator
$$
R^t=(B^t+A^tR^0)(A^{t*}+B^{t*}R^0)^{-1}
$$
belongs to $W_{\varphi^t}$.

3. The following relation is satisfied:
\begin{equation}
||i\frac{R^{t+\delta t}-R^t}{\delta t} -
(Z^t+(L+Y^t)R^t + R^t(Y^{t*}+L) + R^tZ^{t*}R^t)||_2
\rightarrow_{\delta t \rightarrow 0} 0
\end{equation}

}

{\bf Proof.}

1. To prove the first statement, notice that eq.(30) implies that the
matrix (29) is boundedly invertable:
$$
\left(\begin{array}{cc}  A^t & B^t \\ B^{t*} & A^{t*} \end{array}\right)^{-1}
=
 \sum_{n=0}^{\infty}
\left(\begin{array}{cc}  A^t_{-n} & B^t_{-n} \\ B^{t*}_{-n} & A^{t*}_{-n}
\end{array}\right)
\left(\begin{array}{cc}  e^{iLt} & 0 \\ 0 & e^{-iLt} \end{array}\right)
$$
where  $A^t_{-n},B^t_{-n}$ obey the following recursive
relations:
$$
\left(\begin{array}{cc}  A^t_{-n} & B^t_{-n} \\ B^{t*}_{-n} & A^{t*}_{-n}
\end{array}\right)
=i\int_0^t d\tau
\left(\begin{array}{cc}  A^{\tau}_{-n+1} & B^{\tau}_{-n+1} \\
B^{\tau *}_{-n+1} & A^{\tau *}_{-n+1}\end{array}\right)
\left(\begin{array}{cc}  Y^{\tau}_i & Z^{\tau}_i \\
-Z^{\tau *}_i & -Y^{\tau *}_i \end{array}\right)
$$
which imply that
\begin{equation}
\left(\begin{array}{cc}  A^t & B^t \\ B^{t*} & A^{t*} \end{array}\right)^{-1}
=
\left(\begin{array}{cc}  A^{t+} & -B^{tT} \\ -B^{t+} & A^{tT}
 \end{array}\right)
\end{equation}
This means that the matrix (29) is a matrix of a canonical transformation [8].
This
 implies [8] that the operator
$A^t$ is boundedly invertable. As
$$
E-((A^t)^{-1}B^t)
((A^t)^{-1}B^t)^* = (A^t)^{-1} (A^t)^{*-1} >0,
$$
the operator
$(A^{t*})^{-1}B^{t*}$
has the norm lesser than 1. Therefore, the operator
$$
(A^{t*}+B^{t*}R^0)^{-1}=
(E+(A^{t*})^{-1}B^{t*}R^0)^{-1}
(A^{t*})^{-1}
$$
is bounded, since $||R^0||=1$ and
$||(A^{t*})^{-1}B^{t*}||<1.$
The first statement of lemma 7 is proved.

2.
Let us check the property $R^t \in W_2^{\infty}({\bf R}^{2\nu}).$
It follows from the Cauchy-Schwarz-Bunyakovskii inequality that
$$
||\Delta^M R^t \Delta^K||^2_2 = Tr (\Delta^K R^{t*} \Delta^{2M} R^t
\Delta^K) \le ||\Delta^{2M} R^t||_2
||\Delta^{2K} R^t||_2 $$
The first statement of lemma 7 and lemma 5 imply that $$
||\Delta^{2K} R^t||_2  \le ||\Delta^{2K} (B^t+A^tR^0)||_2
||(A^{t*}+B^{t*}R^0)^{-1}|| < \infty.  $$
Therefore, $ R^{t} \in W_2^{\infty}({\bf R}^{2\nu}).$

Let us check that $R^t\varphi^{t*}=-\varphi^t$. As the set of functions
\begin{equation}
u^t=i\varphi^t, v^t=-i\varphi^{t*}
\end{equation}
is a solution to eq.(7), one has
$$
\varphi^{t*}=A^{t*}\varphi^{0*}-B^{t*}\varphi^0=
(A^{t*}+B^{t*}R^0)\varphi^{0*}.
$$
Therefore,
$$
R^t\varphi^{t*}=(B^t+A^tR^0)\varphi^{0*}
=
B^{t}\varphi^{0*}-A^{t}\varphi^0=-\varphi^0.
$$
Eq.(24) is proved.

Let us prove that the operator with the kernel (25) has the norm lesser than
1. It is sufficient to prove that
$  (\xi,(E-R^{t+}R^t)\xi) \ge 0$, $(\xi,(E-R^{t+}R^t)\xi) = 0$
       if and only if $\xi=\lambda \varphi^{t*}.$
We have $$
  (\xi,(E-R^{t+}R^t)\xi) = ( (A^{t*}+B^{t*}R^0)^{-1}\xi,
  (E-R^{0+}R^0) (A^{t*}+B^{t*}R^0)^{-1}\xi).
$$ This quantity is non-negative and equals to zero if and only if
$ (A^{t*}+B^{t*}R^0)^{-1} \xi=\lambda \varphi^{0*}$,
i.e. $\xi=\lambda \varphi^{t*}$.  The second statement is proved.

3. The proof of statement 3 is by straightforward substitution. Lemma 7
is proved.

{\bf Remark.} As the quantity
$\hbar(Z^t+(L+Y^t)R^t + R^t(Y^{t*}+L) + R^tZ^{t*}R^t)$
coincides with the right-hand
 side of eq.(23), lemma 4 is a corollary of lemma 7.

{\bf 3.} Let us give an asymptotic formula for the $N$-particle wave
function being a solution to eq.(4). Let
$\varphi \in W_2^{\infty}({\bf R}^{\nu}), R \in W_{\varphi}.$
Denote by $\Phi_R$ the following element of ${\cal F}_{\varphi}$:
\begin{equation}
\Phi_{R,2n}(x_1,...,x_{2n})=\frac{1}{2^n n! \sqrt{(2n)!}}
\sum_{1<i_1 \ne...\ne i_{2n} < 2n} M(x_{i_1},x_{i_2})...M(x_{i_{2n-1}},
x_{i_{2n}}) ,
\end{equation}
$$
\Phi_{R,2n+1}=0, n=0,1,2,3,....
$$
where $M$ has the form (25). The propery (19) is satisfied for $\Phi_R$
because of eq.(24). It can be shown by making use of the second quantization
technique [8] that the property $||M||<1$ implies that the series (18)
converges. Therefore, $\Phi_R \in {\cal F}_{\varphi}$.

Consider the solution to eq.(4) that satisfies the initial condition
$$
\Psi_N^0=K^N_{\varphi^0}\Phi_{R^0}=
$$
\begin{equation}
\sum_{l=0}^{[N/2]} \frac{1}{(2N)^l l!}
\sum_{1\le i_1 \ne ...\ne i_{2l} \le N} M^0(x_{i_1},x_{i_2})...
M^0(x_{i_{2l-1}},x_{i_{2l}}) \prod_{i \ne i_s} \varphi^0(x_i),
\end{equation}
where
$$
M^0(x,y)=R^0(x,y)+\varphi^0(x)\varphi^0(y),R^0 \in W_{\varphi^0}.
$$
Consider also the solution $\varphi^t$ to eq.(5) which is equal to $\varphi^0$
at initial time moment and the solution $R^t$ to the Cauchy problem for
 eq.(23).Consider the functions:
\begin{equation}
S^t=\int^t_0 dt \left[i\int dx \varphi^{t*}(x)\frac{d}{dt} \varphi^t(x)-
H_0(\varphi^{t*},\varphi^t)\right],
\end{equation}
where
$$
H_0(\varphi^*,\varphi)=\frac{1}{\hbar} \int dx \varphi^*(x) \left(-
\frac{\hbar^2}{2m}\Delta + U(x) \right) \varphi(x) +
$$
$$
\frac{1}{2\hbar} \int dx dy V(x,y) |\varphi(x)|^2 |\varphi(y)|^2;
$$
\begin{equation}
c^t=\exp\left(-\frac{i}{2\hbar} \int^t_0 d\tau \int dx dy V(x,y)
\varphi^{\tau *} (x)
\varphi^{\tau *} (y) R^{\tau}(x,y)\right).
\end{equation}

{\bf Theorem 2.} {\it The following formula is satisfied;}
$$
||\Psi_N^t - c^te^{iNS^t} K^N_{\varphi^t} \Phi_{R^t}||
\rightarrow_{N\rightarrow\infty} 0.
$$

This theorem is a corollary of the more general statement to be proved in
section 6. It follows from theorem 2 and lemma 1 that one can use the
$N$-particle wave function
$$
 c^te^{iNS^t} (K^N_{\varphi^t} \Phi_{R^t})(x_1,...,x_N) =
  c^t e^{iNS^t}
\sum_{l=0}^{[N/2]} \frac{1}{(2N)^l l!}
$$
\begin{equation}
\times\sum_{1\le i_1 \ne ...\ne i_{2l} \le N} M^t(x_{i_1},x_{i_2})...
M^t(x_{i_{2l-1}},x_{i_{2l}}) \prod_{i \ne i_s} \varphi^t(x_i)
\end{equation}
$$
(M^t(x,y)=R^t(x,y)+\varphi^t(x)\varphi^t(y))
$$
instead of the exact wave function $\Psi^t_N$ in order to find limits as
$N \rightarrow \infty$ of mean values of the observables uniformly bounded
with respect to $N$. We see that not the form of the product of
one-particle wave functions but the more complicated form (40) of the
$N$-particle wave function conserves under time evolution. The
product (13) is a partial case of the wave function (40) which is
realized when
\begin{equation} R^t(x,y)=-\varphi^t(x) \varphi^t(y).
\end{equation}
Therefore, one can make use of theorem 2 for finding an
approximate solution to the Cauchy problem for eq.(4) with the initial
condition (13). As the solutioin to eq.(23) does not, in general, have
the form (41), the asymptotic solution to this problem has the form
(40) with $M^t \ne 0$. This implies that eq.(9) is not satisfied
because of corollary
3 from lemma 3. Thus, theorem 1 is a corollary of theorem 2.

\section{Heuristic derivation of the asymptotic formula}

In this section we consider a heuristic method to derive eq.(40). This
method is analogous to the procedure of section 2 which is based on
shifting the solution to the Hartree equation and allows us to conclude that
the product of one-particle wave functions is not an asymptotic solution
to eq.(4).

As the chaos property (2) for correlation functions is satisfied for the
 $N$-particle wave function (20) for arbitrary $g \in {\cal F}_{\varphi}$,
as well as for the function $\Psi_N^t$ [3], it is reasonable to look for
 the asymptotic expression for $\Psi_N^t$ in the following form:
\begin{equation}
\Psi_N^t=e^{iNS^t} K^N_{\varphi^t} g^t, S^t \in {\bf R},
\varphi^t \in L^2({\bf R}^{\nu}), g^t \in {\cal F}_{\varphi}.
\end{equation}
It occurs that the heuristic method to be developed allows us to find the
function $S^t$ up to an additive quantity that does not depend on the
solution to the Hartree equation, while vector $g^t \in {\cal F}_{\varphi^t}$
is defined up to a multiplier $c^t \in {\bf C},|c^t|=1.$

{\bf 1.} The function $S^t$ can be found by the following technique. Consider
the small shift of the function $\varphi$ by the quantity $\chi/N$ of order
$1/N$ in eq.(20). The function $\chi$ can be decomposed into two parts:
\begin{equation}
\chi = \varphi(\varphi,\chi)+\chi^{'},
\end{equation}
where $\chi^{'}$ is orthogonal to $\varphi$. It follows from eq.(20) that
the contribution of shifting by $\chi^{'}$ to formula (20) is small.
We will see that the coefficient of $\chi^{'}$ chould be of order
$1/\sqrt{N}$ for making such shifting of $\varphi$ by $\chi^{'}$ appreciable.
The contribution of the first term of eq.(43) to eq.(20) is as follows:
the $p$-th term of eq.(20) is multiplied by
$$
(1+\frac{1}{N}(\varphi,\chi))^{N-p}.
$$
As $p=O(1)$, all these quantities are approximately equal to
$\exp (\varphi,\chi)$. Because of the remark after lemma 3, norms of terms
of order $p=O(N)$ in eq.(20) are small. Therefore, the following approximate
formula takes place:
\begin{equation}
K^N_{\varphi+\frac{1}{N}\chi} g \simeq e^{(\varphi,\chi)}K^N_{\varphi} g
\end{equation}
This formula can be also proved rigorously [9].

Let us make use of eq.(44) for finding $S^t$. Let us shift the initial
condition for eq.(5)
$
\varphi^0 \rightarrow \varphi^0+\frac{1}{N} \chi^0,
$
so that the solution to eq.(5) will be transformed as
$$
\varphi^t \rightarrow \varphi^t+\frac{1}{N} \chi^t.
$$
The factor $S^t$ depending on the initial condition $\varphi^0$ will be
changed as
$
S^t(\varphi^0) \rightarrow
S^t( \varphi^0+\frac{1}{N} \chi^0).
$
It follows from eq.(44) that
$$
e^{iNS^t(\varphi^0+\frac{1}{N}\chi^0)}e^{(\varphi^t,\chi^t)-
(\varphi^0,\chi^0)} \simeq e^{iNS^t(\varphi^0)}.
$$
Therefore, the variations of $S^t$ and $\varphi^0,\varphi^t$ are related
as follows:
\begin{equation}
\delta S^t=i[(\varphi^t, \delta \varphi^t)
-(\varphi^0, \delta \varphi^0)].
\end{equation}
Eq.(45) determines the function $S^t$ up to a constant that depends on $t$
but does not depend on $\varphi^0$. Let us check that eq.(38) really
satisfies eq.(45). One has:
$$
\delta S^t = \int_0^t dt \int dx
[i(\delta \varphi^t)^*(x) \frac{d}{dt}\varphi^t(x)-
i\frac{d}{dt} (\delta \varphi^t)^*(x) \delta\varphi^t(x)-
(\delta \varphi^t)^*(x) \frac{\delta H_0}{\delta \varphi^{*t}(x)}-
$$
$$
-\delta \varphi^t(x) \frac{\delta H_0}{\delta \varphi^{t}(x)}]
+\int^t_0 dt \int dx i\frac{d}{dt}[\varphi^{t*}(x)\delta \varphi^t(x)].
$$
The first term in this formula vanishes because of the Hartree equation,
while the second term is equal to eq.(45). Thus, the function $S^t$ (38)
can be found by the developed technique up to a function depending only on
$t$ but not on the solution to eq.(5).

{\bf 2.} Let us derive conditions on the function
$g^t \in {\cal F}_{\varphi^t}.$ We are going to consider the
variations of $\varphi^t$ which are of order $1/\sqrt{N}$ and to
obtain formula analogous to eq.(44) for the $N$-particle wave function
\begin{equation}
K^N_{\varphi+\frac{1}{\sqrt{N}}\chi} (g + O(1/\sqrt{N})),
\end{equation}
where $g \in {\cal F}_{\varphi}.$ Notice that the vectors $g+O(1/\sqrt{N})$
belong to different subspaces ${\cal F}_{\varphi+\frac{1}{\sqrt{N}}\chi}$
of the space $\cal F$ and, therefore, should be different in general.

First of all, consider the expression (46) in the case of vacuum vector
$g=\Phi^{(0)}$, i.e. $g_0=1,g_{\alpha}=0,\alpha \ge 1.$ The wave function
(46) is then equal to the product of one-particle wave functions
\begin{equation}
(\varphi(x_1)+\frac{1}{\sqrt{N}}\chi(x_1))
..(\varphi(x_N)+\frac{1}{\sqrt{N}}\chi(x_N))
\end{equation}
which has been considered in section 2. Examine the case $(\varphi,\chi)=0.$
It follows from eq.(20) that formula (47) can be written as
$$
K^N_{\varphi} C_{\chi},
$$
where $C_{\chi}$ is the following element of ${\cal F}_{\varphi}$:
\begin{equation}
C_{\chi,n}(x_1,...,x_N)=\frac{1}{\sqrt{n!}}\chi(x_1)...\chi(x_n).
\end{equation}
When $(\varphi,\chi)\ne 0$, one has
\begin{equation}
K^N_{\varphi+\frac{1}{\sqrt{N}}\chi} \Phi^{(0)} =
(1+\frac{1}{\sqrt{N}} (\varphi,\chi))^N K^N_{\varphi} C_{\chi^{'}},
\end{equation}
where $\chi^{'}$ is expressed from eq.(43).

Consider now the following case of vector $g$ in formula (46):
$
g+O(1/\sqrt{N})=C_{f_N}, f_N=f+O(1/\sqrt{N}).
$
As
\begin{equation}
(\varphi+\frac{1}{\sqrt{N}}\chi,f_N)=0,
\end{equation}
one has
\begin{equation}
K^N_{\varphi+\frac{1}{\sqrt{N}}\chi}C_{f_N}=
K^N_{\varphi+\frac{1}{\sqrt{N}}(\chi+f_N)}\Phi^{(0)}=
(1+\frac{1}{\sqrt{N}}(\varphi,\chi+f_N))^N K^N_{\varphi}
C_{\chi^{'}+f_N-\varphi(\varphi,f_N)}.
\end{equation}
It follows from eq.(50) that
$$
(\varphi,f_N)=-(\chi,f_N)/\sqrt{N}=O(1/\sqrt{N}),
$$
$$
(1+\frac{1}{\sqrt{N}}(\varphi,\chi+f_N))^N \simeq
(1+\frac{1}{\sqrt{N}}(\varphi,\chi))^N e^{-(\chi,f)}.
$$
Since $(\varphi,f)=0$, one has
\begin{equation}
K^N_{\varphi+\frac{1}{\sqrt{N}}\chi}(C_{f}+O(1/\sqrt{N}))=
const K^N_{\varphi} (e^{-(\chi^{'},f)}C_{f+\chi^{'}}+O(1/\sqrt{N})),
\end{equation}
where constant in the right-hand side of eq.(52) does not depend on $f$. As
the vectors (48) make up a full system of the vectors in the space
${\cal F}_{\varphi}$ [8],  eq.(52) can be used for constructing an
approximation for eq.(46).

It is convenient to introduce creation and annihilation operators  [8]
in Fock space $\cal F.$ These operators are the following:
$$
(a^+(x)g)_n(x_1,...,x_n)=\frac{1}{\sqrt{n}} \sum_{i=1}^n \delta(x-x_i)
g_{n-1}(x_1,...,x_{i-1},x_{i+1},...,x_n),
$$
$$
(a^-(x)g)_{n-1}(x_1,...,x_{n-1})=\sqrt{n}g_n(x,x_1,...,x_{n-1}).
$$
Notice that the operators
$$
a^+[\lambda]=\int a^+(x)\lambda(x) dx,
a^-[\lambda]=\int a^-(x)\lambda^*(x) dx,
$$
transform the subspace ${\cal F}_{\varphi}$ into
 ${\cal F}_{\varphi}$ if and only if
 $(\varphi,\lambda)=0.$ Consider the operator
\begin{equation}
V_{\varphi,\chi}=\exp(\int dx (a^+_{\varphi}(x)\chi(x)-
 a^-_{\varphi}(x)\chi^*(x)),
\end{equation}
where
$$
 a^-_{\varphi}(x)=a^-(x)-\varphi(x)a^-[\varphi],
 a^+_{\varphi}(x)=a^+(x)-\varphi^*(x)a^+[\varphi].
$$
The canonical commutation relations
$$
[a^{\pm}(x),a^{\pm}(y)]=0, [a^-(x),a^+(y)]=\delta (x-y)
$$
and the representation of $C_{f}$ through the creation operators
$$
C_f=\exp(a^+[f]) \Phi^{(0)}
$$
imply that the operator (53) transforms the vector (48) as follows:
\begin{equation}
V_{\varphi,\chi}C_f=\exp(-(\chi^{'},f)-\frac{1}{2}(\chi^{'},\chi^{'}))
C_{f+\chi^{'}}.
\end{equation}

The factor $e^{-(\chi^{'},\chi^{'})/2}$ can be involved into the constant
in eq.(52). Therefore, one has
\begin{equation}
K^N_{\varphi+\frac{1}{\sqrt{N}}\chi}(g+O(1/\sqrt{N}))=const
 K^N_{\varphi} (V_{\varphi,\chi}g + O(1/\sqrt{N}))
\end{equation}
for the case $g=C_f$. Since the system of vectors (48) is full [8],
eq.(55) is correct for arbitrary $g$.

{\bf 3.} Let us
find vector $g^t$ entering to eq.(42).
Make use of eq.(55) for finding operator ${\cal W}^t$
transforming $g^0$ into $g^t$, $ g^t={\cal W}^t g^0.$ Suppose that
variation of this operator is also of orde $N^{-1/2}$ as the solution
 to eq.(5) is shifted by the quantity $O(1/\sqrt{N})$. The $N$-particle
 wave functions
 $$
 K^N_{\varphi^t+\frac{1}{\sqrt{N}}\chi^t} {\cal W}^t g + O(1/\sqrt{N})
 $$
 and
 $$
 const (K^N_{\varphi^t} {\cal W}^t V_{\varphi^0,\chi^0} g + O(1/\sqrt{N}))
 $$
 are then asymptotic solutions to eq.(4) if $\varphi^t$ is a solution to
 eq.(5) and the set $u^t=\chi^t, v^t=\chi^{t*}$ is a solution to the
 system (7). As the asymptotic solutions coincide at initial time moment
 because of eq.(55), the same property should be satisfied at time moment
 $t$. Making use of eq.(55), one finds that
\begin{equation}
V_{\varphi^t,\chi^t} {\cal W}^t = const {\cal W}^t V_{\varphi^0,\chi^0}.
\end{equation}
It is eq.(56) that allows us to find operator ${\cal W}^t$ up to a
multiplicative factor. It follows from eq.(27) that eq.(56) is satisfied
if and only if
$$
\int dx [a^+_{\varphi^t}(x)(A^tu^0+B^tv^0)(x)-
a^-_{\varphi^t}(x)(B^{t*}u^0+A^{t*}v^0)(x)]
{\cal W}^t=
$$
\begin{equation}
{\cal W}^t
\int dx [a^+_{\varphi^0}(x)u^0(x)-
a^-_{\varphi^0}(x)v^0(x)].
\end{equation}
Eq.(57) is a straightforward corollary of eq.(56) when
$u^0=\chi^0,v^0=\chi^{0*}.$
In order to prove eq.(56) at arbitrary $u^0,v^0$,
one should write eq.(56) for
$u^0=\chi^0e^{ia},v^0=\chi^{0*}e^{-ia}$
and integrate it over $a$ with the weight $e^{\pm ia}.$

Eq.(57) shows us that the operator ${\cal W}^t$ corresponds to a linear
canonical transformation of the creation and annihilation operators [8].
 It has been shown in [8] that ${\cal W}^t$ is defined up to
 a multiplicative factor.
 In order to find the vector ${\cal W}^t \Phi_{R^0}$ and to
show a role of the Riccati-type equation (23), it is convenient to introduce
a notion of complex germ analogous to [10,6].

{\bf 4.} Let $R \in W_{\varphi}$. Consider the following subspace ${\cal G}_R$
of the space $L^2({\bf R}^{\nu})\times L^2({\bf R}^{\nu}):$
$$
{\cal G}_R=\{ (v,u)|u=Rv \}.
$$

{\bf Definition 2.}
{\it
A subspace ${\cal G}_R$ will be referred to as a complex germ corresponding
to $R \in W_{\varphi}.$
}

The vector $\Phi_R$ satisfies the following interesting property.

{\bf Lemma 8.}
{\it
1. Let $g\in {\cal F}_{\varphi}$ be such vector that
\begin{equation}
\int dx [a^+_{\varphi}(x)u(x)-
a^-_{\varphi}(x)v(x)]g=0
\end{equation}
for  any $(v,u)\in {\cal G}_R$. Then $g=c\Phi_R$.

2. The vector $g=\Phi_R \in {\cal F}_{\varphi}$ satisfies eq.(58).
}

To prove this lemma, one can make use of the definition of creation and
annihilation operators, rewrite eq.(58) in terms of components $g_n$ of
$g$ and find them by induction to be equal to expression (36) up to a
multiplier.

Lemma 7 implies the following statement.

{\bf Lemma 9.}
$$
{\cal G}_{R^t}=\{ (B^{t*}u^0+A^{t*}v^0, A^t u^0+B^t v^0)|(v^0,u^0) \in
 {\cal G}_{R^0} \}
$$

{\bf Corollary.}
\begin{equation}
{\cal W}^t\Phi_{R^0}=c^t\Phi_{R^t}
\end{equation}
for some constant $c^t \in {\bf C}$.

Namely, consider the vector ${\cal W}^t \Phi_{R^0}.$ It follows from eq.(57)
that this vector satisfies the condition (58) for $(v^t,u^t) \in
{\cal G}_{R^t}$, and, therefore, is equal to $c\Phi_{R^t}$.

Therefore, theorem 2 is heuristically justified. Notice that the developed
method can be applied also to the simpler case of the
complex germ approximation
for quantum mechanics as $\hbar \rightarrow 0$, see appendix A for more
details.

Notice also that eq.(57) allows us to construct another asymptotic solutions
to eq.(4) with the help of the complex germ creation operators:
\begin{equation}
\Lambda_{u,v}^{\varphi+}
=\int dx
 [a^+_{\varphi}(x)v^*(x)-
a^-_{\varphi}(x)u^*(x)].
\end{equation}
It follows from eq.(57) that
\begin{equation}
{\cal W}^t
\Lambda_{u_1^0,v_1^0}^{\varphi^0+}...
\Lambda_{u_k^0,v_k^0}^{\varphi^0+}\Phi_{R^0}=c^t
\Lambda_{u_1^t,v_1^t}^{\varphi^t+}...
\Lambda_{u_k^t,v_k^t}^{\varphi^t+}\Phi_{R^t}.
\end{equation}
As any element of the space ${\cal F}_{\varphi}$ can be presented as a
linear (maybe, infinite) combination of vectors (61), eq.(61) allows
us to reconstruct the operator ${\cal W}^t$ and to find asymptotic solutions
$$
c^te^{iNS^t} K^N_{\varphi^t} {\cal W}^t g
$$
to eq.(4).

The approach developed here can be used for constructing asymptotic solutions
 to
equations of a more general form than (4). Let us formulate the corresponding
theorem.

\section{General case : formulation of the theorem}

Let $\cal X$ be a measure space, $H_N$ be a self-adjoint operator in
$L^2({\cal X}^N)$ of the form
\begin{equation}
H_N=NH_0^N+H_1^N+...+N^{1-k}H^N_k,
\end{equation}
where
$$
(H^N_l\Psi)(x_1,...,x_N)=\sum_{p=1}^{P_0}\frac{1}{N^pp!}
\sum_{1\le i_1\ne ... \ne i_p \le N}
\int dy_{i_1}...dy_{i_p}
{\bf H}_l^{(p)}(x_{i_1},...,x_{i_p};y_{i_1},...,y_{i_p})
$$
\begin{equation}
\times
\Psi(x_1,...,x_{i_1-1},y_{i_1},x_{i_1+1},...,x_{i_p-1},y_{i_p},x_{i_p+1},...,
x_N),
\end{equation}
where ${\bf H}^{(p)}_l \equiv
H^{(p)}_l$ is a kernel of the operator acting in $L^2({\cal
X}^p)$ which is symmetric separately over $x_{i_1},...,x_{i_p}$ and
 over $y_{i_1},...,y_{i_p}$ and $$
H_l^{(p)*}(x_1,...,x_p;y_1,...,y_p)=H_l^{(p)}(y_1,...,y_p;x_1,...,x_p).  $$
Analogously to section 3, denote by $\cal F$ the space of sets
$(g_0,g_1(x_1),g_2(x_1,x_2),...)$ of functions $g_n:{\cal X}^n \rightarrow
{\bf C}$ which are symmetric with respect to $x_i \in {\cal X}$, belong
to the spaces $L^2({\cal X}^n)$ and satisfy eq.(18). By
${\cal F}_{\varphi} \subset {\cal F}$ we denote such subspace of $\cal F$ that
consists of all the elements of $\cal F$ which satisfy eq.(19). The
multiparticle canonical operator of the form (20) is denoted by
$K^N_{\varphi}:{\cal F}_{\varphi} \rightarrow L^2({\cal X}^N)$. Denote by
$\Phi_R \in {\cal F}_{\varphi}$ eq.(36).

Let $\Psi_N^t$ be a solution to the Cauchy problem
$$
i\frac{d}{dt}\Psi_N^t = H_N\Psi_N^t,
$$
\begin{equation}
\end{equation}
$$
\Psi_N^0=K^N_{\varphi^0}
\Lambda^{\varphi^0+}_{u_1^0,v_1^0}...
\Lambda^{\varphi^0+}_{u_k^0,v_k^0}\Phi_{R^0},
$$
where $\Lambda^+$ has the form (60).

Denote
$$
{\bf H}_l(\varphi^*,\varphi)=\sum_{p=1}^{P_0} \frac{1}{p!} \int dx_1...dx_p
dy_1...dy_p {\bf H}_l^{(p)}(x_1,...,x_p;y_1,...,y_p)
$$
\begin{equation}
\times\varphi^*(x_1)...
\varphi^*(x_p)\varphi(y_1)...\varphi(y_p), \varphi \in L^2({\cal X}),
\end{equation}
$H_l(\varphi^*,\varphi)\equiv {\bf H}_l(\varphi^*,\varphi).$
Let $\varphi^t$ be a solution to the Cauchy problem for the following equation:
\begin{equation}
i\frac{d}{dt}\varphi^t(x)=\frac{\delta H_0(\varphi^{t*},\varphi^t)}
{\delta \varphi^*(x)},
\end{equation}
such that $\int dx |\varphi^0(x)|^2 =1$, $R^t$ satisfy the equation:
$$
i\frac{d}{dt}R^t(x,y)=
\frac{\delta^2 H_0}{\delta \varphi^*(x) \delta\varphi^*(y)}
+\int dx^{'}
\frac{\delta^2 H_0}{\delta \varphi^*(x) \delta\varphi(x^{'})}
R^t(x^{'},y)
$$
\begin{equation}
+\int dy^{'}
R^t(x,y^{'})
\frac{\delta^2 H_0}{\delta \varphi(y^{'}) \delta\varphi^*(y)}
+\int dx^{'} dy^{'}
R^t(x,x^{'})
\frac{\delta^2 H_0}{\delta \varphi(x^{'}) \delta\varphi(y^{'})}
R^t(y^{'},y),
\end{equation}
where arguments $\varphi^{t*},\varphi^t$ of the function $H_0$ are omitted,
$u_i^t,v_i^t$ satisfy the following system:
$$
i\frac{d}{dt} u_i^t(x)=\int dy \left(
\frac{\delta^2 H_0}{\delta \varphi^*(x) \delta\varphi(y)}u_i^t(y)
+
\frac{\delta^2 H_0}{\delta \varphi^*(x) \delta\varphi^*(y)}v_i^t(y) \right),
$$
\begin{equation}
\end{equation}
$$
-i\frac{d}{dt} v_i^t(x)=\int dy \left(
\frac{\delta^2 H_0}{\delta \varphi(x) \delta\varphi(y)}u_i^t(y)
+
\frac{\delta^2 H_0}{\delta \varphi(x) \delta\varphi^*(y)}v_i^t(y) \right).
$$
By $S^t$ we denote expression (38), while
\begin{equation}
c^t=\exp\left(-i\int_0^t d\tau \left[\frac{1}{2}\int dxdy
\frac{\delta^2 H_0}{\delta \varphi(x) \delta\varphi(y)}R^t(x,y)+H_1
\right]\right),
\end{equation}
where arguments $\varphi^{t*},\varphi^t$ of the functions $H_0,H_1$ are
also omitted.

Let the following functions of $x_1,...,x_m,z_1,...,z_{s-2j-i}$:
$$
\int dx_{m+1}...dx_p dy_1...dy_p \varphi^{t*}(x_{m+1})...\varphi^{t*}(x_p)
H^{(p)}_l(x_1,...,x_p;y_1,...,y_p)
$$
$$
\times\varphi^t(y_{s+1})...\varphi^t(y_p)
R^t(y_1,y_2)...R^t(y_{2j-1},y_{2j})
R^t(y_{2j+1},z_1)...R^t(y_{s-i},z_{s-2j-i})
$$
\begin{equation}
\times\chi^t_{J_1}(y_{s-i+1})...\chi^t_{J_i}(y_s)
\end{equation}
belong to $L^2({\cal X}^{m-s-2j-i})$, where $J_1,...,J_i \in {1,...,k}$,
$\chi_J=v_J^*-Ru_J^*$.

{\bf Theorem 3.}
{\it The following relation takes place:}
$$
||\Psi_N^t-c^te^{iNS^t} K^N_{\varphi^t}
\Lambda^{\varphi^t+}_{u_1^t,v_1^t}...
\Lambda^{\varphi^t+}_{u_k^t,v_k^t}\Phi_{R^t}||
 \rightarrow_{N\rightarrow\infty} 0.
$$
{\bf Remarks.}

1. For some special choice of $H^{(p)}$, eq.(4) is a partial case of
 eq.(64), the Hartree equation (5) is a partial case of eq.(66),
 eq.(23) is the analog of eq.(67), eq.(68) is a variation system (7).
 Therefore, theorems 1 and 2 are corollaries of theorem 3.

2. An asymptotic formula being approximately equal to the wave function
$\Psi_N^t$ accurate to $O(N^{-L/2})$ for arbitrary $L>0$ is to be presented
in section 7.

3. The requirement for the functions (70) to belong to $L^2$ means that the
$N$-particle wave function (61) belongs to the domain of an operator $H_N$.
This requirement can be easily checked for the case of bounded operators
$H_l^N$, while for general case one should prove this independently.
For the case of conditions of theorem 2, the square integrability of
eq.(70) is a corollary of the properties of the solutions to the Hartree
equation (5) and to Riccati equation (23); some other cases are presented in
[11-13].

4. We have used different notations, ${\bf H}_l^{(p)}$ and $H_l^{(p)}$, for
the same quantity. This has been done in order to simplify formulas to appear
 in section 8. In that sectoion we will denote by ${\bf H}_l^{(p)}$ an
 operator in $l^2$, while its eigenvalue will be denoted as $H_l^{(p)}$.
 The same remark is correct for the notations ${\bf H}_l$ and
 $H_l$.

\section{Proof of the theorem}

{\bf 1.} Let us consider the more convenient representation for the operator
$H_N$ that allows us to find a commutation rule between $H_N$ and
multiparticle canonical operator $K^N_{\varphi}$.

Consider the space $L^2({\cal X}^N)$ as a subspace ${\cal F}^N$ of the
Fock space $\cal F$ of the form
$$
{\cal F}^N=\{ g\in {\cal F}| g_{\alpha}=0,\alpha\ne N \}
$$
Consider the operator
$$
H^N_l=\sum_{p=1}^{P_0} \frac{1}{N^pp!} \int dx_1...dx_p
dy_1...dy_p {\bf H}_l^{(p)}(x_1,...,x_p;y_1,...,y_p)
$$
\begin{equation}
\times a^+(x_1)...
a^+(x_p)a^-(y_1)...a^-(y_p).
\end{equation}

{\bf Lemma 10.} {\it
The operator (71) transforms ${\cal F}^N$ into ${\cal F}^N$ and coincide
on the subspace ${\cal F}^N$ with the operator (63).
}

The proof is by making use of the definition of creation and annihilation
operators.

As the operator $H_N$ is expressed through the operators $a^{\pm}$, it is
sufficient to find their commutation rules with the multiparticle
canonical operator.

{\bf Lemma 11.}
{\it The following relations are satisfied:}
$$
\frac{a_{\varphi}^+(x)a^-[\varphi]}{\sqrt{N}} K^N_{\varphi}=
K^N_{\varphi} a^+_{\varphi}(x),
$$
\begin{equation}
\end{equation}
$$
a^-_{\varphi}(x)K^N_{\varphi}=\frac{a^-[\varphi]}{\sqrt{N}}K^N_{\varphi}
a^-_{\varphi}(x).
$$
The proof is by making use of the following expression for the element
$K^N_{\varphi} g \in {\cal F}^N \subset {\cal F}$:
$$
K_{\varphi}^N g =\sum_{p=0}^N \frac{1}{\sqrt{p!}} \int dx_1...dx_p
g_p(x_1,...,x_p)
$$
\begin{equation}
\times \frac{a^+_{\varphi}(x_1)a^-[\varphi]}{\sqrt{N}}...
 \frac{a^+_{\varphi}(x_p)a^-[\varphi]}{\sqrt{N}}
 \frac{a^+[\varphi]^N}{\sqrt{N!}}  \Phi^{(0)}
\end{equation}
and of the commutation relations between operators $a^{\pm}(x)$.

Consider now the operator $\frac{d}{dt}K^N_{\varphi}$.

{\bf Lemma 12.}
{\it  The following relation takes place;}
$$
\frac{d}{dt} K^N_{\varphi} =K^N_{\varphi} (\frac{d}{dt} + a^+[\varphi]
a^-_{\varphi}[\frac{d\varphi}{dt}]+
(\varphi,\frac{d}{dt}\varphi)(N-\hat{n})
$$
\begin{equation}
-
(1-\frac{\hat{n}}{N})\sqrt{N}
a^-_{\varphi}[\frac{d\varphi}{dt}]
+\sqrt{N}
a^+_{\varphi}[\frac{d\varphi}{dt}]),
\end{equation}
{\it where an index $t$ on $\varphi$ is omitted,}
$$
a^+_{\varphi}[\chi]=a^+[\chi]-a^+[\varphi](\varphi,\chi),
a^-_{\varphi}[\chi]=a^-[\chi]-a^-[\varphi](\chi,\varphi),
\hat{n}=\int dx a^+_{\varphi}(x) a^-_{\varphi}(x).
$$
{\bf Proof.} Consider the $N$-particle wave function
$$
\frac{d}{dt} K^N_{\varphi} g = \frac{d}{dt} \sum_{p=0}^N
\sqrt{\frac{N!}{N^pp!}}\frac{1}{(N-p)!}
$$
\begin{equation}
\times
 \int dx_1...dx_p g_p(x_1,...,x_p)
a^+(x_1)...a^+(x_p) a^+[\varphi]^{N-p} \Phi^{(0)}, g \in {\cal F}_{\varphi}.
\end{equation}
When one takes a derivative with respect to $t$, there will be two terms:
one of them contains $\frac{d}{dt}g_p$, another contains
$\frac{d}{dt}a^+[\varphi]$. Consider the first term. The function
$\frac{dg}{dt}$ can be decomposed into two parts:
$$
\frac{d}{dt} g_p(x_1,...,x_p) = -\sum_{i=1}^p \varphi(x_i)
 \int \frac{d \varphi^*(y_i)}{dt} g_p(x_1,...,x_{i-1},y_i,x_{i+1},...,x_p)
 dy_i
 $$
 $$
 + \frac{d^{'}}{dt} g_p(x_1,...,x_p).
$$
The second part being equal to
$$
\frac{d^{'}}{dt} g = (\frac{d}{dt} + a^+[\varphi] a^-_{\varphi}
[\frac{d}{dt}\varphi]) g
$$
belongs to ${\cal F}_{\varphi}$, since the property $g \in {\cal F}_{\varphi}$
(19) conserves under time evolution, and contribute to eq.(75) as
$K^N_{\varphi} \frac{d^{'}}{dt} g$. The first part gives rise to the
following contribution to the expression (75):
$$
-\sum_{p=0}^N \sqrt{\frac{N!}{N^pp!}}\frac{1}{(N-p)!} \sqrt{p}
\int (a^-_{\varphi}[\frac{d}{dt} \varphi]g)_{p-1}(x_1,...,x_{p-1})
$$
$$
\times
a^+(x_1)...a^+(x_{p-1}) dx_1...dx_{p-1} (a^+[\varphi])^{N-p+1} \Phi^{(0)}
$$
which can be presented as
$$
-\frac{1}{\sqrt{N}}K^N_{\varphi}(N-\hat{n})
a^-_{\varphi}\left[\frac{d\varphi}{dt}\right].
$$
Consider now the terms containing $\frac{d}{dt}a^+[\varphi]$. They
contribute to eq.(75) as follows:  $$
\sum_{p=0}^N \sqrt{\frac{N!}{N^pp!}}\frac{1}{(N-p)!} \int
g_{p}(x_1,...,x_{p}) a^+(x_1)...a^+(x_{p}) dx_1...dx_{p}
$$
$$
\times
(N-p) a^+[\frac{d}{dt} \varphi] (a^+[\varphi])^{N-p} \Phi^{(0)}.
$$ According
 to the definition of $a^{\pm}_{\varphi}$, this expression is equal to
$$
K^N_{\varphi} \left( (\varphi,\frac{d}{dt}\varphi) (N-\hat{n}) +
\sqrt{N} a^+_{\varphi}[\frac{d}{dt}\varphi]\right).
$$
Combining all the terms, we obtain eq.(74). Lemma 12 is proved.

{\bf Corollary.}
\begin{equation}
(i\frac{d}{dt}-H_N)
e^{iNS^t}K^N_{\varphi^t}=
e^{iNS^t}K^N_{\varphi^t}
\left(i\frac{d}{dt}+ia^+[\varphi^t]a^-_{\varphi^t}[\frac{d}{dt}\varphi^t]
-{\cal H}_N^{'}\right),
\end{equation}
where
$$
{\cal H}_N^{'}=N\frac{d}{dt}S^t - iN (\varphi^t,\frac{d}{dt}\varphi^t)
(1-\frac{\hat{n}}{N}) - i \sqrt{N} a^+_{\varphi^t}[\frac{d}{dt}\varphi^t]
+i\sqrt{N}(1-\frac{\hat{n}}{N}) a^-_{\varphi^t}[\frac{d}{dt}\varphi^t]
$$
$$
+
\sum_{l=0}^k N^{1-l} \sum_{p=1}^{P_0} \sum_{m,s=0}^{p}
\frac{p!}{m!(p-m)!s!(p-s)!}\frac{1}{N^{\frac{m+s}{2}}}
 \int dx_1...dx_p dy_1...dy_p
$$
$$
\times
\varphi^{t*}(x_{m+1})...\varphi^{t*}(x_p)
{\bf H}^{(p)}_l(x_1,...,x_p;y_1,...y_p)
\varphi^t(y_{s+1})...\varphi^t(y_p)
a_{\varphi^t}^+(x_1)...a_{\varphi^t}^+(x_m)
$$
\begin{equation}
\times (1-\hat{n}/N)(1-(\hat{n}+1)/N)...(1-(\hat{n}+p-m-1)/N)
a_{\varphi^t}^-(y_1)...a_{\varphi^t}^-(y_s).
\end{equation}
{\bf Remark.}
The corollary implies that the operator ${\cal H}_N^{'}$ is a product of
$N$ by a polynomial in $N^{-1/2}$:
\begin{equation}
{\cal H}_N^{'}=NH_0^{'}+N^{1/2}H_1^{'}+H_2^{'}+...+N^{1-K/2}H^{'}_K
\end{equation}
for some $K$. The coefficients $H_0^{'},H_1^{'},H_2^{'}$ can be presented
as follows:
\begin{equation}
H^{'}_0={\bf H}_0(\varphi^{t*},\varphi^t)+\frac{d}{dt}S^t-
i(\varphi^t,\frac{d}{dt}\varphi^t),
\end{equation}
\begin{equation}
H^{'}_1 =\int dx \left[
a^+_{\varphi^t}(x) \left(
\frac{\delta {\bf H}_0}{\delta \varphi^*(x)}-i\frac{d}{dt}
\varphi^t(x)) +
a^-_{\varphi^t}(x) (\frac{\delta {\bf H}_0}{\delta \varphi(x)}-i\frac{d}{dt}
\varphi^{t*}(x)\right) \right],
\end{equation}
$$
H_2^{'}=\hat{n} \int dx \varphi^{t*}(x) \left(i\frac{d}{dt}\varphi^t(x)-
\frac{\delta {\bf H}_0}{\delta \varphi^*(x)}\right)
 + {\bf H}_1(\varphi^{t*},\varphi^t)
$$
$$
-\frac{1}{2}\int dx dy \varphi^t(x) \varphi^t(y)
 \frac{\delta^2 {\bf H}_0}{\delta \varphi(x) \delta \varphi(y)}
+\int dx dy [\frac{1}{2}
a^+_{\varphi^t}(x)
 \frac{\delta^2 {\bf H}_0}{\delta \varphi^*(x) \delta \varphi^*(y)}
a^+_{\varphi^t}(y)
$$
\begin{equation}
+
a^+_{\varphi^t}(x)
 \frac{\delta^2 {\bf H}_0}{\delta \varphi^*(x) \delta \varphi(y)}
a^-_{\varphi^t}(y)
+\frac{1}{2}
a^-_{\varphi^t}(x)
 \frac{\delta^2 {\bf H}_0}{\delta \varphi(x) \delta \varphi(y)}
a^-_{\varphi^t}(y)],
\end{equation}
where the arguments $\varphi^{t*},\varphi^t$ of the functional ${\bf H}_0$
are omitted.
One can notice that eqs.(38),(66) provide the nullification of
$H^{'}_0,H^{'}_1$.

{\bf 2.}
Let us prove that
\begin{equation}
||(i\frac{d}{dt}-H_N)
e^{iNS^t}K^N_{\varphi^t}g^t|| \rightarrow_{N\rightarrow\infty} 0,
\end{equation}
where
$$
g^t=c^t
\Lambda^{\varphi^t+}_{u_1^t,v_1^t}...
\Lambda^{\varphi^t+}_{u_k^t,v_k^t}\Phi_{R^t}.
$$
{\bf Lemma 13.}
$$
||H_k^{'} g^t||<\infty, k=\overline{0,K}.
$$

The proof is by making use of commutation relations between creation and
annihilation operators, formula
\begin{equation}
\Phi_R=\exp\left(
\frac{1}{2}\int dx dy a_{\varphi}^+(x) M(x,y) a_{\varphi}^+(y)\right)
\Phi^{(0)}
\end{equation}
and the condition of the square integrability of the functions (70).

As $H_0^{'}=H_1^{'}=0$, for checking eq.(82) it is sufficient to prove
the following lemma.

{\bf Lemma 14.}
$$
(i\frac{d}{dt} + i a^+[\varphi]a^-_{\varphi}[\frac{d}{dt}\varphi]-
H_2^{'}) g^t=0.
$$

{\bf Proof.}
1. It follows from straigthforward calculaton that
\begin{equation}
[i\frac{d}{dt} + i a^+[\varphi^t]a^-_{\varphi^t}[\frac{d}{dt}\varphi^t]-
H_2^{'},
\Lambda^{\varphi^t+}_{u^t,v^t}]f^t=0, f^t \in {\cal F}_{\varphi^t}
\end{equation}
if and only if
$$
i\frac{d}{dt} v^{t'*}(x)-\int dy \left(
\frac{\delta^2 H_0}{\delta \varphi^*(x) \delta\varphi(y)}v^{t'*}(y)
+
\frac{\delta^2 H_0}{\delta \varphi^*(x) \delta\varphi^*(y)}u^{t'*}(y) \right)
=\alpha^t \varphi^t(x),
$$
$$
i\frac{d}{dt} u^{t'*}(x)+\int dy \left(
\frac{\delta^2 H_0}{\delta \varphi(x) \delta\varphi(y)}v^{t'*}(y)
+
\frac{\delta^2 H_0}{\delta \varphi(x) \delta\varphi^*(y)}u^{t'*}(y) \right)
=\beta^t \varphi^{t*}(x),
$$
where
$$
u^{t'}=u^t-\varphi^t(\varphi^t,u^t),
v^{t'}=v^t-\varphi^{t*}(\varphi^{t*},v^t),
$$
$\alpha^t,\beta^t$ are some complex functions. This system for
$u^{t'},v^{t'}$, as well as eq.(84),
is satisfied    when $(u^t,v^t)$ satisfies eq.(68).

2. It is sufficient then to prove lemma for the case $k=0$. From eq.(83) one
has:
$$
(i\frac{d}{dt} + i a^+[\varphi]a^-_{\varphi}[\frac{d}{dt}\varphi])
c^t\Phi_{R^t}=i\frac{dc^t}{dt}\Phi_{R^t} +
\frac{i}{2}c^t \int dxdy a^+_{\varphi^t}(x) \frac{dM^t}{dt}(x,y)
a^-_{\varphi^t}(y) \Phi_{R^t},
$$
$$
H_2^{'}c^t\Phi_{R^t}=c^t (H_1+\frac{1}{2}\int dx dy (M^t(x,y)-\varphi^t(x)
\varphi^t(y))
\frac{\delta^2H_0}{\delta\varphi(x)\delta\varphi(y)})
$$
$$
+\frac{c^t}{2} \int dx dy
 a^+_{\varphi^t}(x) a^+_{\varphi^t}(y)[
\frac{\delta^2H_0}{\delta\varphi^*(x)\delta\varphi^*(y)}+
\int dz
\frac{\delta^2H_0}{\delta\varphi^*(x)\delta\varphi(z)}M^t(z,y)
$$
$$
+\int dz M^t(z,x)
\frac{\delta^2H_0}{\delta\varphi(z)\delta\varphi^*(y)}+
\int dz dz^{'} M^t(z,x) M^t(z^{'},y)
\frac{\delta^2H_0}{\delta\varphi(z)\delta\varphi(z^{'})} ]\Phi_{R^t}.
$$
Lemma 14 is then proved as a corollary of eqs.(67),(69).

Eq.(82) implies the statement of the theorem,the proof is
analogous to [14]. Consider the quantity
$
e^{iNS^t}K^N_{\varphi^t}g^t -\Psi^t_N
$
being equal to
$$
e^{iNS^t}K^N_{\varphi^t}g^t -\Psi^t_N=
\int_0^t d\tau e^{-iH_N(t-\tau)}(i\frac{d}{d\tau}-H_N)
(e^{iNS^{\tau}}K^N_{\varphi^{\tau}}g^{\tau} -\Psi^{\tau}_N)
$$
The following estimation takes place
$$
||e^{iNS^t}K^N_{\varphi^t}g^t -\Psi^t_N||\le
\int_0^t d\tau ||(i\frac{d}{d\tau}-H_N)e^{iNS^{\tau}}K^N_{\varphi^{\tau}}
g^{\tau}||
\rightarrow_{N\rightarrow\infty} 0.
$$
Theorem 3, as well as theorem 1,2, is proved.

\section{Corrections to the asymptotic formula}

Let us construct now the $N$-particle wave function that approximates the
solution to the Cauchy problem for eq.(64) accurate to $O(N^{-L/2})$
for arbitrary $L>0$. It happens that such wave function has the form:
$$
\Phi_{N,L}^t=e^{iNS^t}K^N_{\varphi^t}
(g^t_0+N^{-1/2}g^t_1+...+N^{-(L-1)/2}g^t_{L-1}),
$$
where vectors $g_m^t \in {\cal F}_{\varphi^t}$ are of the form
$$
g^t_m=\sum_{n=0}^{n_0(m)} \frac{1}{\sqrt{n!}}\int dx_1...dx_n
g^t_{m,n}(x_1,...,x_n)
a^+_{\varphi^t}(x_1)...a^+_{\varphi^t}(x_n)\Phi_{R^t},
$$
$n_0(m)$ is a finite quantity, functions $g_{m,n}(x_1,...,x_n)$ are to be
defined by induction. Let $g^t_{m,n}$ be defined for $m<l$ and all functions
$$
\int dx_{q+1}...dx_p dy_1...dy_p \varphi^{t*}(x_{q+1})...\varphi^{t*}(x_p)
{\bf H}^{(p)}_r(x_1,...,x_p;y_1,...,y_p)
$$
$$
\times\varphi^t(y_{s+1})...\varphi^t(y_p)
R^t(y_1,y_2)...R^t(y_{2j-1},y_{2j})
R^t(y_{2j+1},z_1)...R^t(y_{s-n},z_{s-2j-n})
$$
\begin{equation}
\times g^t_{m,n}(y_{s-n+1},...,y_s)
\end{equation}
belong to $L^2({\cal X}^{m-s-2j-n})$
for $m<l$. Define the functions
 $\chi^t_{l,n}(x_1,...,x_n)$
 being symmetric with respect to
$x_i$ and obeying the requirement
$
\int dx_1 \varphi^{t*}(x_1)
\chi^t_{l,n}(x_1,...,x_n)=0
$
by the following relation:
$$
\sum_n \frac{1}{\sqrt{n!}}
\int dx_1...dx_n \chi^t_{l,n}(x_1,...,x_n) a^+_{\varphi^t}(x_1)
..a^+_{\varphi^t}(x_n)\Phi_{R^t}=
$$
\begin{equation}
=H_3^{'}g^t_{l-1}+...+H_K^{'}g^t_{l-K},
\end{equation}
where all $g_{n-i}^t=0$ by definition
as $n<i$. Note that definition (86) is correct and
$\chi^t_{l,n}=0$ for sufficiently large $n$, the sum in the left-hand
side of eq.(86) is then finite.
When $l=0$, let $\chi^t_{0,n}=0$ by definition.

Let $g^t_{l,n}$ be a solution to the Cauchy problem for the following
equation:
$$
i(\frac{d}{dt}-\frac{d\ln c^t}{dt})
 g^t_{l,n}(x_1,...,x_n)=\sum_{k=1}^n \int dy_k
[\frac{\delta^2H_0}{\delta \varphi^*(x_k)\delta\varphi(y_k)}+
$$
$$
\int dz_k R^t(x_k,z_k)
\frac{\delta^2H_0}{\delta \varphi(z_k)\delta\varphi(y_k)}]
g^t_{l,n}(x_1,...,x_{k-1},y_k,x_{k+1},...,x_n)
+\frac{(m+1)(m+2)}{2}
$$
\begin{equation}
\times \int dy_1 dy_2
\frac{\delta^2H_0}{\delta \varphi(y_1)\delta\varphi(y_2)}
g^t_{l,n+2}(y_1,y_2,x_1,...,x_n)
+\chi^t_{l,n}(x_1,...,x_n),
\end{equation}
where $c^t$ has the form (69).
Let $\Psi^t_N$ be a solution to eq.(64) that satisfies the
initial condition
$$
\Psi_N^t=K^N_{\varphi^0}(g^0_0+N^{-1/2}g^0_1+...+N^{-(L-1)/2}g^0_{L-1}).
$$
{\bf Theorem 4.}
{\it Under the conditions of theorem 3}
$$
||\Psi^t_N-\Phi_{N,L}^t|| =O(N^{-L/2}).
$$

{\bf Remarks.}

1. The requirement for the functions (79) to be square integrable is
 analogous to the same requirement for the function (68) and means that
 the asympyotic solution $\Phi_{N,L}^t$ belongs to the domain of an
 operator $H_N$.

2.For the case of eq.(4), the solvability of eq.(81) and the square
integrability of eq.(79) can be proved in a way analogous to the proof of
solvability of eqs.(7) and (23). When $H_N$ is a bounded operator,
such statements can be also proved.

{\bf Proof.}
 It is sufficient to show that
 $$
||(i\frac{d}{dt}-H_N)\Phi^t_{N,L}||=O(N^{-L/2}),
 $$
i.e.
\begin{equation}
(i\frac{d}{dt}+ia^+[\varphi^t]a^-_{\varphi^t}[\frac{d}{dt}\varphi]-
H^{'}_2)g^t_m=H^{'}_3g^t_{m-1}+...+H^{'}_Kg^t_{m-K}.
\end{equation}
The proof of this relation is straightforwrd. Theorem 4 is proved.
\section{Some aspects of problems with operator-valued symbols}

We have seen that the discussed method of constructing asymptotic solutions
can be applied to equations of the form (64). However, one can be interested
in the problem of generalization of the considered approach to the case of the
set of such equations. Investigation of it is very important when one
considers the quantum mechanical system consisting of two subsystems (one
of them is the examined system of $N$ bose-particles, another subsystem
interacts with the first one), some examples are to be discussed in section 9.
Notice that some of the results to be obtained in this section and in
section 9 can be also derived by making use of the technique analogous
to the derivation of the Ehrenfest theorem in ordinary quantum mechanics
[20], see appendix B for more details.
Let us now consider the specification of the form of the set of equations
which is to be approximately solved.

Let ${\cal X}$ be a measure space. Denote by
$l^2 \otimes L^2({\cal X}^N)$
the Hilbert space of sets of complex functions $\Psi_I(x_1,...,x_N),
I=\overline{1,\infty}, x_1,...,x_N \in {\cal X}$ such that
$\Psi_I \in L^2({\cal X}^N)$ and
$\sum_{I=1}^{\infty} \int dx_1...dx_N |\Psi_I(x_1,...,x_N)|^2 <
\infty.$ Let $H_N$ be a self-adjoint operator in $l^2 \otimes
L^2({\cal X}^N)$ of the form (62); operators $H^N_l$ have the form
(63), where $\Psi \in l^2 \otimes L^2({\cal X}^N)$, while ${\bf
H}^{(p)}_l(x_1,...,x_p;y_1,...,y_p)$ being operators in $l^2$ with
matrices
\newline
$H^{(p)}_{l,IJ}(x_1,...,x_p;y_1,...,y_p)$
($I,J=\overline{1,\infty}$) are kernels of the following operators
$\hat{H}^{(p)}_l$ in $l^2 \otimes L^2({\cal X}^N)$:
 $$
(\hat{H}^{(p)}_l \Psi)_I(x_1,...,x_p)=\int dy_1...dy_p
H^{(p)}_{l,IJ}(x_1,...,x_p;y_1,...,y_p)\Psi_J(y_1,...,y_p),
$$
$$
\Psi \in
l^2 \otimes L^2({\cal X}^N),
$$
which are also required to be self-adjoint. Denote by
$l^2 \otimes {\cal F}$
the space of sets ${\bf g}=(g_{0,I},g_{1,I}(x_1),g_{2,I}(x_1,x_2),...)$
of functions $g_{n,I}:{\cal X}^n \rightarrow {\bf C}$,
 $I=\overline{1,\infty}$, which are symmetric with respect to
 $x_i \in {\cal X}$, belong to the spaces $L^2({\cal X}^n)$ and satisfy
 the following condition analogous to eq.(18):
 $$
 \sum_{I=1}^{\infty} \sum_{n=0}^{\infty} \int dx_1...dx_n
 |g_{n,I}(x_1,...,x_n)|^2 <\infty.
 $$
 By $l^2 \otimes {\cal F}_{\varphi}
 \subset l^2 \otimes {\cal F}$, where $\varphi \in L^2({\cal X})$, we
 denote, analogously to the case of section 5, such subspace of
 $l^2 \otimes {\cal F}$
that consists of all the elements of
 $l^2 \otimes {\cal F}$
that satisfy the condition
$$
\int dx_1 \varphi^*(x_1) g_{n,I}(x_1,...,x_n)=0.
$$
Denote by
${\bf K}_{\varphi}^N :
l^2 \otimes {\cal F}_{\varphi} \rightarrow
l^2 \otimes L^2({\cal X}^N)$
the following analog of the multiparticle canonical operator (20):
$$
({\bf K}_{\varphi}^N {\bf g})_I(x_1,...,x_N)=
(K_{\varphi}^N g_I)(x_1,...,x_N),
$$
where ${\bf g}\in l^2 \otimes {\cal F}$, while $g_I$
is the element of ${\cal F}$ of the form
$(g_{0,I},g_{1,I}(x_1),g_{2,I}(x_1,x_2),...)$ at fixed $I$.

We are going to find approximate solutions to eq.(64) for the case of
the operator-valued function ${\bf H}^{(p)}_l$ and
$\Psi_N^t \in l^2 \otimes L^2({\cal X})$ by the technique analogous to the
method discussed in sections 5-7. These asymptotic solutions are to be
looked for in the following form:
\begin{equation}
\Psi_N^t = e^{iNS^t} {\bf K}^N_{\varphi^t} ({\bf g}_0^t+N^{-1/2}
{\bf g}_1^t +...),
\end{equation}
where ${\bf g}_i^t \in l^2 \otimes {\cal F}_{\varphi^t}$.
We will formulate the theorem in section 9, while in this section we are
to find ${\bf g}_i^t$ heuristically. One should substitute the expression
(89) to eq.(64) and make use of the commutation rule between operators
${\bf K}_{\varphi^t}^N$ and $(id/dt-H_N)$.

{\bf Lemma 14.} {\it The following relation is satisfied:}
\begin{equation}
(i\frac{d}{dt} - H_N)e^{iNS^t}{\bf K}^N_{\varphi^t} =
e^{iNS^t}{\bf K}^N_{\varphi^t} (iD_t - {\cal H}_N^{'}),
\end{equation}
{\it where} ${\cal H}_N^{'}$ {\it has the form (77) and}
$$
D_t=d/dt + a^+[\varphi^t]a^-_{\varphi^t}[\frac{d}{dt}\varphi^t].
$$

The proof of this lemma for the case of the operator-valued function
${\bf H}_l^{(p)}$ is analogous to the proof of eq.(76) for the case of
sections 5-7.

By ${\bf H}_l(\varphi^*,\varphi)$ we denote the operator in $l^2$ of the form
(65). The operator ${\cal H}^{'}_N$ in $l^2 \otimes {\cal F}_{\varphi}$
is then written in a form (78), where operators $H_0^{'},H_1^{'},H_2^{'}$
have the form (79),(80),(81). For the simplicity, the operators in $l^2$
of the form like $\lambda E$, where $\lambda$ is a number, are denoted by
$\lambda$.

One can notice that one should choose ${\bf g}_0^t,{\bf g}_1^t,...$ in
such a way that
\begin{equation}
(-NH_0^{'}-N^{1/2}H_1^{'}+iD_t-H_2^{'}-N^{-1/2}H_3^{'}-...)
({\bf g}^t_0+N^{-1/2}{\bf g}_1^t+...) = O(N^{-L/2})
\end{equation}
when the asymptotics accurate to $O(N^{-L/2})$ is looked for.

An interesting feature of the operator-valued case is that the operators
$H_0^{'},H_1^{'}$ cannot be set to zero by varying $\varphi^t$, since
the operator ${\bf H}_0(\varphi^*,\varphi)$ is not, in general,
equal to $H_0(\varphi^*,\varphi)$. Therefore, in order to provide satisfaction
of the relations like eq.(82), one cannot choose ${\bf g}^t$ to be
independent on $N$, ${\bf g}^t$ must be choosen as
${\bf g}^t={\bf g}_0^t+N^{-1/2}{\bf g}_1^t+N^{-1}{\bf g}_2^t.$

Let us consider the recursive relations for ${\bf g}_i^t$, which are
derivable from eq.(91). First of all, consider the term of order $O(N)$
in eq.(91) which has the form $H_0^{'}{\bf g}_0^t=0$. It follows from eq.(79)
that $H_0^{'}$ can be presented as ${\bf H}_0(\varphi^{t*},\varphi^t)-
\lambda^t$ for some number $\lambda^t$. Therefore, ${\bf g}_0^t$ should
be chosen as ${\bf g}_0^t=\zeta \otimes {\bf g}_0^t$, i.e.
$$
g_{0,n,I}^t(x_1,...,x_n)=\zeta_I g_{0,n}^t (x_1,...,x_n),
$$
where $\zeta \in l^2$ is the eigenvector of the operator
${\bf H}_0(\varphi^{t*},\varphi^t)$ acting in $l^2$.

Let $H_0(\varphi^{t*},\varphi^t)$ be eigenvalue of the operator
${\bf H}_0(\varphi^{t*},\varphi^t)$ and smooth function of $t$,
${\Pi}(\varphi^{t*},\varphi^t)$ be projector on the corresponding
eigenspace, so that
\begin{equation}
({\bf H}_0(\varphi^{t*},\varphi^t)-
H_0(\varphi^{t*},\varphi^t))
{\Pi}(\varphi^{t*},\varphi^t)=0.
\end{equation}
The term of order $O(N)$ vanishes then, if $S^t$ has the form (38).

Suppose that this eigenspace is one-dimensional. The method under
consideration can be also applied in analogous way to the case of
finite and  $\varphi^t$-independent dimensionality of the eigenspace.
The case of terms intersection, ehen this dimensionality depends on
$\varphi$, requires the more careful treatment.

Assume that $H_0$ is an isolated point of the spectrum of the operator
${\bf H}_0$ in $l^2$, so that there exists a unique operator $R$ such
that
$$
R\Pi=0, ({\bf H}_0-H_0)R(1-\Pi)=1-\Pi.
$$
We will denote this operator $R$ as
$$
({\bf H}_0-H_0)^{-1}(1-\Pi) = R,
$$
the arguments ${\varphi}^{t*},\varphi^t$ of the operator $\Pi$
and functional $H_0$ are omitted.

To each operator $A$ in $l^2$ with the matrix $A_{IJ}$ we assign the operator
in $l^2 \otimes {\cal F}_{\varphi}$ that transforms the vector with
components $g_{n,I}(x_1,...,x_n)$ into the vector
\newline
$\sum_{J=1}^{\infty} A_{IJ} g_{n,J}(x_1,...,x_n).$ This operator in
$l^2 \otimes {\cal F}_{\varphi}$ will be also denoted by the same
symbol,$A$.

Let us consider other terms of eq.(91). It is convenient to present
${\bf g}_i^t$ as
$$
{\bf g}^t_i ={\bf g}^{t\parallel}_i + {\bf g}^{t\perp}_i,
$$
where
$$
{\bf g}_i^{t\parallel}=\Pi {\bf g}_i^t,
{\bf g}_i^{t\perp}=(1-\Pi) {\bf g}_i^t.
$$
It follows from eq.(38) that $H_0^{'}{\bf g}^{t\parallel}_i =0.$
Therefore, eq.(91) can be written as
\begin{equation}
({\bf H}_0 - H_0) {\bf g}^{t\perp}_m+H_1^{'}{\bf g}_{m-1}^t+
(H_2^{'}-iD_t){\bf g}^t_{m-2}+...+H_K^{'} {\bf g}^t_{m-K} =0.
\end{equation}
The vector ${\bf g}_m^{t\perp}$
is determined in a unique fashion  from this relation if and only
if
\begin{equation}
\Pi (H_1^{'}{\bf g}_{m-1}^t+
(H_2^{'}-iD_t){\bf g}^t_{m-2}+...+H_K^{'} {\bf g}^t_{m-K}) =0,
\end{equation}
since $\Pi ({\bf H}_0 -H_0)=0$ (because ${\bf H}_0$ is a self-adjoint
operator).

When $m=1$, eq.(94) implies that $\Pi H_1^{'} {\bf g}_0^t=0$. It follows
from eq.(80) that one should require that
$$
\Pi \left(i\frac{d}{dt}\varphi^t(x) -
 \frac{\delta {\bf H}_0}{\delta\varphi^*(x)} \right) \zeta=0,
$$
i.e.
\begin{equation}
\Pi \left(i\frac{d}{dt}\varphi^t(x) -
 \frac{\delta {\bf H}_0}{\delta\varphi^*(x)} \right) \Pi=0.
\end{equation}
This equation is the analog of Hartree equation for the operator-valued case.

Furthermore, one can find ${\bf g}^{t\perp}_{m-1}$ from eq.(93), substitute
it to eq.(94), make use of  formula (95) implying that
$\Pi H_1^{'}{\bf g}_{m-1}^{t\parallel}=0$ and obtain the equation for
${\bf g}^t_{m-2}$. In order to find its solution, one can first find
${\bf g}^{t\perp}_{m-2}$ from eq.(93) and obtain the following equation
for ${\bf g}^{t\parallel}_{m-2}:$
$$
-\Pi H_1^{'} ({\bf H}_0-H_0)^{-1}(1-\Pi)
(H_1^{'}({\bf g}^{t\parallel}_{m-2}+{\bf g}^{t\perp}_{m-2})
+(H_2^{'}-D_t)g^t_{m-3}+H_3^{'}g^t_{m-3}+...)
$$
\begin{equation}
+\Pi ((H_2^{'}-iD_t)({\bf g}^{t\parallel}_{m-2}+{\bf g}^{t\perp}_{m-2})
+...) =0,
\end{equation}
where ${\bf g}^{t\perp}_{m-2}$ are found from eq.(93). Therefore, one can look
for the quantities ${\bf g}_m^{t\perp}$ and ${\bf g}_m^{t\parallel}$
by induction. Let ${\bf g}_0^{t\perp},{\bf g}_0^{t\parallel},...,
{\bf g}^{t\perp}_{m-3},{\bf g}^{t\parallel}_{m-3}$ be already found.
Then one should find ${\bf g}_{m-2}^{t\perp}$ from eq.(93) and
${\bf g}_{m-2}^{t\parallel}$ from eq.(96). The equation for
${\bf g}_{m-2}^{t\parallel}$
has the form like
\begin{equation}
\Pi (iD_t -H_2^{'}+ H_1^{'}({\bf H}_0-H_0)^{-1}(1-\Pi)H_1^{'})
{\bf g}_{m-2}^{t\parallel}=
{\chi}_{m-2}^{t\parallel}
\end{equation}
for some right-hand side
${\chi}_{m-2}^{t\parallel}$
such that
$(1-\Pi ){\chi}_{m-2}^{t\parallel}=0$.
Eq.(97) is analogous then to eq.(88).

Let us simplify our main equations, (95) and (97). As
$\Pi ({\bf H}_0-H_0)=({\bf H}_0-H_0) \Pi =0$, one has
$$
0=\frac{\delta}{\delta \varphi^*(x)}(\Pi({\bf H}_0-H_0)\Pi )=
\Pi\left(\frac{\delta ({\bf H}_0-H_0)}{\delta \varphi^*(x)}\right)\Pi .
$$
Therefore, eq.(95) takes the form of eq.(66), while
$$
H_1^{'}=\int dx \left[
a^+(x)\left(\frac{\delta ({\bf H}_0-H_0)}{\delta \varphi^*(x)}\right)
+a^-(x)\left(\frac{\delta ({\bf H}_0-H_0)}{\delta \varphi(x)}\right)
\right]
$$
In order to simplify eq.(97), denote by $\zeta^t \in l^2$, $||\zeta^t||=1$,
the eigenvector of ${\bf H}_0(\varphi^{t*},\varphi^t)$. One has
\begin{equation}
{\bf g}_{m-2}^{t\parallel}=\zeta^t \otimes g^t_{m-2},
{\chi}_{m-2}^{t\parallel}=\zeta^t \otimes \chi^t_{m-2}.
\end{equation}
Making use of the relations like
$$
\Pi \frac{\delta ({\bf H}_0-H_0)}{\delta \varphi(x)}=
-\frac{\delta \Pi}{\delta \varphi(x)}({\bf H}_0-H_0),
$$
$$
2\frac{\delta\Pi}{\delta \varphi(y)}
 \frac{\delta ({\bf H}_0-H_0)}{\delta \varphi(x)}\Pi +
 \Pi  \frac{\delta^2 ({\bf H}_0-H_0)}{\delta \varphi(x)\delta\varphi(y)}
\Pi=0,
$$
$$
\Pi \frac{\delta ({\bf H}_0-H_0)}{\delta \varphi(y)}({\bf H}_0-H_0)^{-1}
(1-\Pi)
 \frac{\delta ({\bf H}_0-H_0)}{\delta \varphi(x)}\Pi
 $$
$$
=-\frac{\delta \Pi}{\delta \varphi(y)}(1-\Pi)
 \frac{\delta ({\bf H}_0-H_0)}{\delta \varphi(x)}\Pi =
-\frac{\delta \Pi}{\delta \varphi(y)}
 \frac{\delta ({\bf H}_0-H_0)}{\delta \varphi(x)}\Pi
$$
and commutation relations between operators $a^{\pm}(x)$,
one obtains that eq.(97) takes the following form:
$$
[iD_t+\gamma^t-\int dx dy
[\frac{1}{2}a_{\varphi^t}^+(x)
\frac{\delta^2H_0}{\delta\varphi^*(x)\delta\varphi^*(y)}
a_{\varphi^t}^+(y)
$$
\begin{equation}
+a_{\varphi^t}^+(x)
\frac{\delta^2H_0}{\delta\varphi^*(x)\delta\varphi(y)}
a_{\varphi^t}^-(y)
+\frac{1}{2}a_{\varphi^t}^-(x)
\frac{\delta^2H_0}{\delta\varphi(x)\delta\varphi(y)}
a_{\varphi^t}^-(y)]g^t_{m-2}=\chi^t_{m-2},
\end{equation}
where $\gamma^t$ is a number of the form:
$$
\gamma^t=i\left(\zeta^t,\frac{d\zeta^t}{dt}\right) +
(\zeta^t,\frac{1}{2} \int dxdy [\varphi(x)
\frac{\delta^2{\bf H}_0}{\delta\varphi(x)\delta\varphi(y)}
\varphi(y)-\varphi(x)
\frac{\delta^2({\bf H}_0-H_0)}{\delta\varphi(x)\delta\varphi^*(y)}
\varphi^*(y)]
$$
\begin{equation}
-{\bf H}_1) \zeta^t)- \int dx
\left(\frac{\delta\zeta^t}{\delta\varphi(x)},
\frac{\delta^2({\bf H}_0-H_0)}{\delta\varphi^*(x)} \zeta^t\right).
\end{equation}
Notice that eq.(92) has been already solved in sections 6,7. Therefore,
the functions ${\bf g}_m^t$ are found.

Note that the leading asymptotics has the following form:
\begin{equation}
\Psi_N^t=\tilde{c}^te^{iNS^t}{\bf K}_{\varphi^t}^N
(\zeta^t \otimes
\Lambda^{\varphi^t+}_{u_1^t,v_1^t}...
\Lambda^{\varphi^t+}_{u_k^t,v_k^t}\Phi_{R^t}) + O(1/\sqrt{N}).
\end{equation}
It is remarkable that the functions $\varphi^t,u_k^t,v_k^t,R^t$ obeys the
equations coinciding with the equations obtained in sections 5-7 for the
case of a single Schr\"{o}dinger-like equation, not for a set. The only
difference with the considered case is that the quantity (100) arises in
the left-hand side of eq.(99), so that the phase factor $\tilde{c}^t$ differs
from the factor $c^t$ obtained in sections 5-7.This confirms the heuristic
arguments of section 4 that predict the form of the asymptotics by making use
of the equation for $\varphi^t$ only.

Consider the terms of the additional factor $\gamma^t$ in more details.
The first term was studied in ref.[14] for the case of semiclassical
approximation (the Maslov canonical operator with real phase) for quantum
mechanics. When one considers adiabatic perturbation theory, this term known
as Berry phase [18] also arises.

Consider the last term of eq.(100). One can formally rewrite it as
\begin{equation}
\frac{1}{2} \int dx
(\zeta^t,
\frac{\delta^2({\bf H}_0-H_0)}{\delta\varphi(x)\delta\varphi^*(x)} \zeta^t).
\end{equation}
We will study some examples for which the form (100) is correct, while
 the expression (102) contains divergences to be eliminated. Therefore, we
 will use eq.(100).

Discuss now the problem of chaos conservation for the operator-valued case.
One can study different eigenvalues and eigenvectors of the operator
${\bf H}_0$ and obtain asymptotic solutions being superpositions of the
formulas like eq.(101):
\begin{equation}
\sum_{J=1}^{\infty} \tilde{c}^t_Je^{iNS^t}{\bf K}_{\varphi^t_J}^N
(\zeta^t_J \otimes
\Lambda^{\varphi^t+}_{u_{1,J}^t,v_{1,J}^t}...
\Lambda^{\varphi^t+}_{u_{k_J,J}^t,v_{k_J,J}^t}\Phi_{R^t_J}) + O(1/\sqrt{N}).
\end{equation}
One can consider the quantities analogous to the $k$-particle correlators (1)
$$
{\cal R}^{t}_{k,N}(x_1,...,x_k;y_1,...,y_k)=
$$
\begin{equation}
\sum_{J=1}^{\infty} \int dx_{k+1}...dx_N
\Psi^{t}_{N,J}
(x_1,...,x_k,x_{k+1},...,x_N)\Psi^{t*}_{N,J}(y_1,...,y_k,x_{k+1},...,x_N),
\end{equation}
where $\Psi_N^t\in l^2\otimes L^2({\cal X}^N)$. The correlation functions
(104), as well as the correlators (1), allow us to predict the limits as
$N\rightarrow\infty$ of mean values of the observables of the special form
(3). We can notice that for the element (103) of the space
$l^2 \otimes L^2({\cal X}^N)$ such correlators (104) have limits
$$
{\cal R}^t_{k,N}(x_1,...,x_k;y_1,...,y_k)
 \rightarrow_{N\rightarrow\infty}
 \sum_{J=1}^{\infty} \lambda_J
 \varphi_J^t(x_1)... \varphi_J^t(x_k)
 \varphi_J^{t*}(y_1)... \varphi_J^{t*}(y_k),
$$
where $\lambda_J$ are some numbers.
For example, one can choose the functions $\varphi_I^t$ to be
coinciding at the initial time moment. Since the functions $\varphi_I^t$
obey different Hartree-like equations corresponding to diferent
eigenvalues of ${\bf H}_0$, they will not, in general, coincide at time
 moment $t$. Therefore, the chaos property (2) being satisfied at $t=0$
 does not hold at arbitrary time moment. Therefore, even for the correlation
 functions, the chaos conservation hypothesis fails for the operator-valued
 case.

\section{Operator-valued case: the theorem and some examples}

{\bf 1}. Let $H_0(\varphi^*,\varphi)$ be non-degenerate eigenvalue of the
operator ${\bf H}_0(\varphi^*,\varphi)$, $\varphi^t$ be a solution to eq.(66),
$S^t$ have a form (38). Let ${\bf g}_n$ satisfy recursive relations (93)
and the norm of the vectors $H_m^{'}{\bf g}_n,D_t{\bf g}_n$ be finite.
Consider the solution to the equation
$$
i\frac{d}{dt} \Psi_N^t=H_N \Psi^t_N,
\Psi^t_N \in l^2 \otimes L^2({\cal X}^N)
$$
that satisfies the initial condition
\begin{equation}
\Psi_N^0={\bf K}^N_{\varphi^0} ({\bf g}_0^0 + N^{-1/2}{\bf g}_1^0+...+
N^{-L/2}{\bf g}^0_L).
\end{equation}

{\bf Theorem 5.} {\it The following relation is satisfied:}
$$
||\Psi_N^t-e^{iNs^t}
{\bf K}^N_{\varphi^t} ({\bf g}_0^t + N^{-1/2}{\bf g}_1^t+...+
N^{-L/2}{\bf g}^t_L)||=O(N^{-\frac{L+1}{2}}).
$$
{\bf Remarks}.

1. The initial condition (105) for the Cauchy problem is not arbitrary.
In particular,
 ${\bf g}^0_0$ should be an eigenvector of the operator ${\bf H}_0$.
Moreover, the vectors ${\bf g}^{0\perp}_m$ can be expressed through
${\bf g}_{m-1}^0,...,{\bf g}_0^0$ with the help of eq.(93).
When the initial condition does not satisfy these properties, one can
present it as a superposition of the permissable initial conditions
corresponding to different eigenvalues of ${\bf H}_0$. Therefore, one
should make use of the solutions to different Hartree-like  equations
(66). The chaos property (2) will not conserve then under time evolution.

2. When the initial condition (105) is allowable, one can solve the Cauchy
problem for the recursive relations by induction with the help of the
technique analogous to the previous section: one can first express
${\bf g}_m^{t\perp}$ through ${\bf g}^t_{m-1},...,{\bf g}_0^t$, then one
should find the solution to the Cauchy problem for eq.(97) by using the
substitution (98) and reducing eq.(97) to eq.(99). One can notice that
if the initial condition for ${\bf g}^t_i$ is expressed as a result of
action of a polynomial in creation operators $a^+_{\varphi}(x)$ to
the vector $\Phi_R$, then the function $\chi_{m-2}^t$ is also expressed
in such a way:
$$
\chi_l^t=\sum_n \frac{1}{\sqrt{n!}} \int dx_1...dx_n
\chi_{l,n}^t(x_1,...,x_n) a^+_{\varphi^t}(x_1)...a^+_{\varphi^t}(x_n)
\Phi_{R^t}.
$$
The vector function ${\bf g}^t_{m-2}$ has then the form
$$
{\bf g}_l^t= \exp(i\int_0^t \Gamma^t dt)
\sum_n \frac{1}{\sqrt{n!}} \int dx_1...dx_n
g_{l,n}^t(x_1,...,x_n) a^+_{\varphi^t}(x_1)...a^+_{\varphi^t}(x_n)
\Phi_{R^t},
$$
where  $l=m-2$,
\begin{equation}
\Gamma^t=\gamma^t-\frac{1}{2}\int dx dy
\frac{\delta^2H_0}{\delta\varphi(x)\delta\varphi(y)}  M^t(x,y),
\end{equation}
$\gamma^t$ has the form (100),
while the set of functions $g^t_{l,n}(x_1,...,x_n)$ is a solution to the
Cauchy problem for eq.(87).

3. The proof of  theorem 5 is analogous to the proofs of theorems 3,4.
Nevertheless, one should take into account that the function
$$
{\bf K}^N_{\varphi^t} ({\bf g}_0^t + N^{-1/2}{\bf g}_1^t+...+
N^{-L/2}{\bf g}^t_L).
$$
approximately satisfies eq.(64) accurate to
$O(N^{-\frac{L-1}{2}})$, not to $O(N^{-\frac{L+1}{2}})$.
Therefore, one should substitute to eq.(64) the asymptotic formula with
two additional terms.

4. If the dimensionality of the eigenspace is a constant $D$ more than 1,
one should consider the orthonormal basis $\zeta^t_!,...,\zeta^t_D$ in
this eigenspace and present
 ${\bf g}_{m-2}^{t\parallel}$ as
$$
{\bf g}_{m-2}^{t\parallel}= \sum_{I=1}^D \zeta^t_I \otimes
 {\bf g}^{t(I)}_{m-2}.
$$
The equation for ${\bf g}^{t(I)}_{m-2}$ has then also the form (99), but
$\gamma^t$ should be considered as a matrix $D\times D$ of the form
$$
\gamma^t_{MN}=i\left(\zeta^t_M,\frac{d\zeta_N^t}{dt}\right) +
(\zeta^t_M,\frac{1}{2} \int dxdy [\varphi(x)
\frac{\delta^2{\bf H}_0}{\delta\varphi(x)\delta\varphi(y)}
\varphi(y)-\varphi(x)
\frac{\delta^2({\bf H}_0-H_0)}{\delta\varphi(x)\delta\varphi^*(y)}
\varphi^*(y)]
$$
$$
-{\bf H}_1) \zeta^t_N)- \int dx
\left(\frac{\delta\zeta^t_M}{\delta\varphi(x)},
\frac{\delta({\bf H}_0-H_0)}{\delta\varphi^*(x)} \zeta^t_N\right).
$$

{\bf 2.}
We have considered the case of finding asymptotic solutions to the infinite
set of equations for functions $\Psi_I(x_1,...,x_N),I=\overline{1,\infty}$.
The same method is applicable when one considers the set of $d$ equations for
$d$ functions
 $\Psi_I(x_1,...,x_N),I=\overline{1,d}$,i.e one studies the equation for the
 element of ${\bf R}^d \otimes L^2({\cal X}^N).$

{\bf Example 1.}
 Let us consider the simple example.
 Consider the set of two Schr\"{o}dinger-like equations
$$
i\hbar \frac{\partial}{\partial t}
\left(\begin{array}{c} \Psi^t_{N,1}(x_1,...,x_N)\\
 \Psi^t_{N,2}(x_1,...,x_{N}) \end{array}\right) =
 \sum_{i=1}^N \hbar
\left(\begin{array}{cc} B_{11}(x_i) & B_{12}(x_i) \\
                        B_{21}(x_i) & B_{22}(x_i) \end{array}\right)
\left(\begin{array}{c} \Psi^t_{N,1}(x_1,...,x_N)\\
 \Psi^t_{N,2}(x_1,...,x_{N}) \end{array}\right)
$$
\begin{equation}
+\left[\sum_{i=1}^{N}\left(-\frac{\hbar^2}{2m} \Delta_i + U(x_i)\right)
+\frac{1}{N} \sum_{1 \le i<j \le N} V(x_i,x_j) \right]
\left(\begin{array}{c} \Psi^t_{N,1}(x_1,...,x_N)\\
 \Psi^t_{N,2}(x_1,...,x_{N}) \end{array}\right)
\end{equation}
This set of equations corresponds to the following physical problem. Besides
$N$-particle system, there is a two-level system interacting with $N$
particles. It is the first term in the right-hand side of eq.(107) that
describes this interaction. One can note that the coefficient of this
interaction is of order $O(1)$, contrary to the coefficient of the
particle interaction potential which is $1/N$. Therefore, the term with the
matrix $B$ is to give rise to the additional term in the Hartree equation.

According to the developed technique, consider the matrix
${\bf H}_0(\varphi^*,\varphi)$ corresponding to the operator in ${\bf R}^2$
$$
{\bf H}_0(\varphi^*,\varphi)=H_0^0(\varphi^*,\varphi)
\left(\begin{array}{cc} 1 & 0 \\
                        0 & 1 \end{array}\right)
+ \int dx |\varphi(x)|^2
\left(\begin{array}{cc} B_{11}(x) & B_{12}(x) \\
                        B_{21}(x) & B_{22}(x) \end{array}\right),
$$
where
$$
H_0^0(\varphi^*,\varphi)=\frac{1}{\hbar} \int dx \varphi^*(x) \left(-
\frac{\hbar^2}{2m}\Delta + U(x) \right) \varphi(x) +
$$
$$
+\frac{1}{2\hbar} \int dx dy V(x,y) |\varphi(x)|^2 |\varphi(y)|^2.
$$
The operator ${\bf H}_0$ has two eigenvalues:
$$
H_0^{\pm} (\varphi^*,\varphi)=H_0^0(\varphi^*,\varphi)+
\frac{H_{11}+H_{22}}{2} \pm \beta(\varphi^*,\varphi),
$$
where
$$
H_{IJ}(\varphi^*,\varphi)=\int dx |\varphi(x)|^2 B_{IJ}(x),
\beta=\sqrt{\left(\frac{H_{11}-H_{22}}{2}\right)^2 +H_{12}H_{21}}.
$$
Therefore, there are two Hartree-like equations:
\begin{equation}
i\frac{d}{dt}\varphi^t_{\pm}(x)=\frac{\delta H_0^{\pm}}{\delta \varphi^*(x)}.
\end{equation}
One finds the following asymptotics for the solution to eq.(107):
$$
\Psi_{N,I}^t=
e^{iNS^t_+ + i\int_0^t dt\Gamma^t_+}{\bf K}_{\varphi^t_+}^N
(\zeta^+_I \Lambda^{\varphi_+^t+}_{u_{1,+}^t,v_{1,+}^t}...
\Lambda^{\varphi_+^t+}_{u_{k_+,+}^t,v_{k_+,+}^t}\Phi_{R^t_+}) +
$$
$$
e^{iNS^t_- + i\int_0^t dt \Gamma^t_-}{\bf K}_{\varphi^t_-}^N
(\zeta^-_I \Lambda^{\varphi_-^t+}_{u_{1,-}^t,v_{1,-}^t}...
\Lambda^{\varphi_-^t+}_{u_{k_-,-}^t,v_{k_-,-}^t}\Phi_{R^t_-}) + O(1/\sqrt{N}),
$$
where $S^t_{\pm}$ has the form (38), $\Gamma_{\pm}^t$ has the
form (106), $\varphi^t_{\pm}$ obey eq. (108), $(u_{i,\pm},v_{i,\pm})$
obey the variation system (68), $R^t_{\pm}$ obey eq.(67),
$$
\zeta_1^{\pm}=aH_{12},\zeta_2^{\pm}=a(\frac{H_{22}-H_{11}}{2} \pm \beta),
$$
$$
a^{-2}=2\beta(\frac{H_{22}-H_{11}}{2} \pm \beta).
$$

Let us now consider the problem of divergences in eq. (102). When one formally
calculates the quantity
$
(\zeta,\frac{\delta^2 {\bf H}_0}{\delta \varphi^*(x)\delta \varphi(x)}\zeta),
$
one obtains:
$$
\delta(0) [
\frac{1}{2}(B_{11}(x)+B_{22}(x))
$$
$$
\pm \frac{1}{2\beta}
\left(
(B_{11}(x)-B_{22}(x)) \frac{H_{11}-H_{22}}{2}+H_{21}B_{12}(x)+H_{12}
B_{21}(x)
\right)].
$$
It happens that this infinite quantity is equal to the divergent part of
$
(\zeta,\frac{\delta^2 {H}_0}{\delta \varphi^*(x)\delta \varphi(x)}\zeta).
$
Therefore, the divergences can be eliminated in eq.(102), but they arise
in calculation. In order to avoid arising of infinite quantities,
we have used eq.(100) instead of eq. (102).

{\bf Example 2.}
The developed approach can be also applied to the more interesting case of
eq.(8). To construct asymptotic solutions to eq.(8),one should first
consider the operator
$$
{\bf H}_0=H_0^0 + \frac{1}{\hbar}\left(
-\frac{\hbar^2}{2M}\Delta_y+{\cal U}(y)+\int dz {\cal V}(z,y) |\varphi(z)|^2
\right)
$$
depending on $\varphi$ and find the eigenvalues $H_0^{(I)}(\varphi^*,
\varphi)$
and eigenfunctions $\zeta_I[\varphi^*,\varphi](y)$ of this operator.
Then one should make use of the solutions to the Hartree-like equations (66),
as well as to eqs.(67),(68). The asymptotic solution has then the form (103).
The argumentation on the chaos non-conservation for the correlation
functions which has been presented in the end of section 8 is
also valid for this example.

\section{The problem of chaos conservation and
\newline asymptotic formulas for abstract
Hamiltonian algebras}

{\bf 1.}
We have constructed asymptotic solutions to equations of the type (64) and to
sets of such equations. It was shown that the $L^2({\cal X}^N)$ norm of the
difference between exact and approximate solutions tended to zero as
$N\rightarrow\infty$. Since $\Psi_N^t$ was interpretted as a multiparticle wave
function, lemma 1 told us that its approximation $e^{iNS^t}K^N_{\varphi^t}
g^t$ constructed in sections 5,6 could be used instead of $\Psi_N^t$ for
finding limits as $N\rightarrow\infty$ of mean values of general observables
uniformly bounded with respect to $N$.

On the other hand, there are some other interesting cases. For example, one can
consider the case of classical statistical mechanics when the algebra of
observables is presented not as an operator algebra but as algebra of
real-valued functions on the phase space. The case of quantum statistical
mechanics when states are specified not by wave functions but by density
matrices can  be also studied [13]. Examination of these cases requires one
to formulate analogs of lemma 1.

In this section we are going to generalize lemma 1 for the  case of abstract
Hamiltonian algebras of observables (see, for example, [3,19]) which
involves all the cases considered earlier. Our results will imply, for
example, the conclusion of refs.[11,13] that asymptotic solutions to
Liouville and Wigner equations which are found by the developed technique
should be interpretted not as approximate densities but as approximations
for $N$-particle half-densities which are equal to the square roots of
density functions (matrices) and also obey Liouville (Wigner) equations. We
will introduce the notion of an abstract half-density which generalizes the
notions of refs.[11.13]. We will also consider the equation for it and find
its asymptotic solutions.

These asymptotics are to be expressed through the solutions to eqs.(66),(68).
If we consider $\varphi^t$ and $\varphi^{t*}$ to be independent, the set of
equation (66) and equation conjugated to it will form a Hamiltonian system
playing an important role in constructing tunnel asymptotics [13]. It happens
that for the case under consideration in this section such Hamiltonian
systems can be simplified, so that the number of equations can be cut in half.

{\bf 2.}
{\bf Definition 3.} [3,19]
{\it
 A Hamiltonian algebra $\cal A$ is a set of complex linear space ${\cal A}$
 and mappings
 $\pi:{\cal A}\times {\cal A} \rightarrow {\cal A} $,
 $\lambda:{\cal A}\times {\cal A} \rightarrow {\cal A} $,
 $j:{\cal A} \rightarrow {\cal A} $
 denoted also as $\pi (A,B) \equiv AB$, $\lambda (A,B) \equiv \{ A,B\}$,
 $j(A) \equiv A^+,$
 $A,B \in {\cal A}$, if the following axioms hold:

A1). for any $A,B,C \in {\cal A}, \alpha,\beta\in {\bf C}$

a) $A(\alpha B+\beta C)=\alpha AB +\beta AC,
  \{ A,\alpha B+\beta C \}=\alpha \{ A,B\} +\beta \{ A,C\};$

b) $(A^+)^+=A, (\alpha A + \beta B)^+=\alpha^* A^+ + \beta^* B^+;$

c) $(AB)^+=B^+A^+,\{A,B\}^+=\{ A^+,B^+\};$

d) $\{ A, BC\} = \{ A,B \} C + B \{ A,C\};$

e) $A(BC)=(AB)C;$

f) $\{ A,B\}=-\{ B,A\};  \{\{A,B\},C\}+\{\{B,C\},A\}+\{\{C,A\},B\} = 0;$

A2). there exists such element $I \in {\cal A}$ that
 $AI=IA=A, \{A,I\}=0$ for any $A\in {\cal A}$;

A3). for any $A,B\in {\cal A}$  and some $\hbar\in {\bf R}$
    $ AB-BA=i\hbar \{A,B\}$.
}

{\bf Remark.} The case $\hbar=0$ is also allowable, so that one cannot
write
    $\{A,B\}=\frac{1}{i\hbar} (AB-BA)$.

By $\cal L$ we denote the complex linear space of all linear functionals
$\rho:{\cal A}\rightarrow {\bf C}$. Introduce also the notation
$$
{\cal L}_+=\{\rho\in {\cal L} | \rho(I)=1,\rho(A^+)=(\rho(A))^*,
\rho(A^+A)>0 \forall A\in {\cal L} \}.
$$
Suppose that ${\cal L}_+ \ne \emptyset $.

By ${\cal A}_F \subset {\cal A}$ we denote the set of such elements
$A \in {\cal A}$ that
 $\sup_{\rho\in{\cal L}_+} |\rho(A)| < \infty$.
Consider the following functional $p:{\cal A}_F \rightarrow {\bf R}$:
$$
p(A)=\sup_{\rho\in{\cal L}_+} |\rho(A)|.
$$

{\bf Lemma 15.}
{\it The following relations are satisfied:
$$
p(A+B) \le p(A)+p(B), p(\alpha A)=|\alpha|p(A), A,B \in {\cal A}_F,
\alpha \in {\bf C}.
$$
}

The proof is straightforward.

Let $\{ A\in {\cal A}_F | p(A)\ne 0 \} \ne \emptyset $.
Denote by ${\cal L}_F$ the linear space of such linear functionals
${\cal A}_F \rightarrow {\bf C}$ that
$\sup_{p(A)=1} |\rho(A)| < \infty.$
Inroduce the following norm in ${\cal L}_F$:
$$
||\rho||=
\sup_{p(A)=1} |\rho(A)| , \rho \in {\cal L}_F.
$$

{\bf Lemma 16.}
{\it
The functional $||\cdot||:{\cal L}_F \rightarrow {\bf R}$ satisfies
the following properties:
$$
1) ||\alpha \rho||=|\alpha| ||\rho||,
   ||\rho_1 +\rho_2|| \le ||\rho_1||+||\rho_2||,||\rho||\ge 0,
   \rho,\rho_1,\rho_2\in {\cal L}_F,\alpha\in {\bf C};
$$
$$
2) ||\rho||=0 \Leftrightarrow \rho=0.
$$
}

{\bf Proof.} The proof of the first property is straightforward. Let us
prove the second property. When $\rho=0$, $||\rho||$ is obviously equal to
0. Let $||\rho||=0$ and show that $\rho=0$. One has: $\rho(A)=0$ for
any $A\in{\cal A}_F$ such that $p(A)\ne 0$. It is sufficient to check
that $\rho(B)=0$ when $p(B)=0$. Lemma 15 implies that
$p(A+B)\le p(A)+p(B)=p(A)$,$p(A)\le P(A+B)+p(B)=p(A+B)$. Therefore,
$p(A)=p(A+B)\ne 0$ and $\rho(B)=\rho(A+B)-\rho(A)=0$.
Lemma 16 is proved.

{\bf Remark.} The Hamiltonian algebra $\cal A$ playes the role of the algebra
of observables. The elements of ${\cal L}_+$ specify possible states of the
system. The quantity $\rho(A)$ is then the average value of the observable
$A$ in the state $\rho$. The observables of the form $A^+A$ are usually
called non-negative. The property $\rho(A^+A)\ge 0$ for elements of
${\cal L}_+$ means that average values of non-negative observables are also
non-negative. The functional $p(A)$ has the following physical meaning:
it is the largest possible average value of the observable $\cal A$. The role
of the norm $||\rho||$ is the following: if the quantity $||\rho_1-\rho_2||$
is small, the difference between average values $\rho_1(A)-\rho_2(A)$
of any abservable $A$ such that $p(A)=1$ is also small.

Let us give now some examples of Hamiltonian algebras.

{\bf Example 1.} Denote by ${\cal A}_n^c$ the algebra of smooth functions
$A(p_1,q_1,...,p_n,q_n)$ on ${\bf R}^{2\nu n}$. Define the algebra
operations as follows:
$$
(AB)(X)=A(X)B(X), A^+(X)=A^*(X), X=(p_1,q_1,...,p_n,q_n);
$$
\begin{equation}
\end{equation}
$$
\{ A, B \}(p_1,q_1,...,p_n,q_n)=\sum_{i=1}^n \left(
\frac{\partial A}{\partial q_i}
\frac{\partial B}{\partial p_i}-
\frac{\partial A}{\partial p_i}
\frac{\partial B}{\partial q_i}
\right)
(p_1,q_1,...,p_n,q_n).
$$
It is easy to see that the axiom A1 is satisfied. The element $I(X)\equiv 1$
playes the role of $I$, so that the axiom A2 is also checked. The axiom A3
is satisfied when $\hbar=0$. Therefore, the algebra ${\cal A}_n^c$ is
Hamiltonian. Note that it is the algebra of observables for classical
 satistical mechancs.

{\bf Lemma 17.}
{\it The functional $p(A)$ is the following in the case of
example 1:}
$p(A)=sup_{X\in {\bf R}^{2\nu n}} |A(X)|.$

{\bf Proof.} Let $|A(X)|<C$. Then $2C-A(X)e^{i\phi}-A^+(X)e^{-i\phi} =
B^+B$ for some observable $B$. Therefore, for any $\rho\in {\cal L}_+$
one has $0< \rho(CI)-Re(e^{i\phi}\rho(A))$,i.e. $|\rho(A)|<C$.
Thus,
$
p(A)\le \sup_{X\in {\bf R}^{2\nu n}} |A(X)|.
$
Consider now the element $\rho\in {\cal L}_+$ of the form
$\rho_{X_0}(A)=A(X_0)$. We see that
$p(A)\ge |\rho_{X_0}(A)|=|A(X_0)|$. We prove
lemma 17.

Let us give the examples of elements of ${\cal L}_F$.

A) To any function $\rho$ from $L^1({\bf R}^{2\nu n})$ one can assign te
element $\rho\in {\cal L}_F$ of the form
\begin{equation}
\rho(A)=\int dX \rho(X) A(X), X \in {\bf R}^{2\nu n}.
\end{equation}
One has: $||\rho||=\int dX |\rho(X)|$.

B) The following element of ${\cal L}_F$: $\rho_{X_0}(A)=A(X_0)$ can be
also formally written in the form (110), but $\rho(X)$ is a generalized
function $\rho(X)=\delta(X-X_0)$. One has $||\rho||=1$.

Note that the function $\rho$ playes the role of a probability distribution.

{\bf Example 2.} Let ${\cal A}_n^q$ be algebra of bounded operators
in $L^2({\bf R}^n)$. Let
$$
\pi (A,B)=AB, \lambda (A,B)= \frac{1}{i\hbar} (AB-BA), j(A)=A^+,I=E.
$$
Analogously to the previous example, one can check axioms A1-A3. The
functional $p(A)$ is equal to the ordinary operator norm:
$$
p(A)=||A||=\sup_{||\varphi||=1} (\varphi, A \varphi).
$$
Checking this property is analogous to the proof of lemma 17: one can
present operators $cE-A e^{i\phi} -A^+ e^{-i\phi}$ as $B^+B$.
Note also that elements of ${\cal L}_F$ play the role of density
matrices.

{\bf 3.}
Let us now generalize the notion of half-density to the case of an
abstract Hamiltonian algebra.

{\bf Definition 4.}
{\it
A half-density representation of a Hamiltonian algebra $\cal A$ is a set of
a Hilbert space $\cal H$ and mappings
$\Pi:{\cal A}_F \times {\cal H} \rightarrow {\cal H},$
$\Lambda:{\cal A}_F \times {\cal H} \rightarrow {\cal H},$
denoted also as
$$
\Pi(A,\varphi)\equiv A\varphi \equiv \Pi^A\varphi,
\Lambda(A,\varphi)\equiv \{A,\varphi\} \equiv \Lambda^A\varphi,
A\in {\cal A}_F, \varphi \in {\cal H},
$$
if the following axioms hold:

H1). the mappings $\Pi^A$ and $\Lambda^A$ are linear operators in $\cal H$
which are defined on a common domain ${\cal D} \subset {\cal H}$;

H2).$ \Pi^I=E, \Lambda^I=0; $

H3). the following relations are satisfied for any $A,B \in {\cal A}_F$,
$\varphi,\chi \in {\cal D}$:

a) $\Pi^{\alpha A+\beta B}=\alpha \Pi^A + \beta \Pi^B,
   \Lambda^{\alpha A+\beta B}=\alpha \Lambda^A + \beta\Lambda^B,$

b) $\{ A,B\varphi\}=\{ A,B\} \varphi + B\{A,\varphi\}, $

c) $\Pi^{AB}=\Pi^A\Pi^B,
     \Lambda^{\{A,B\}}=\Lambda^A\Lambda^B-\Lambda^B\Lambda^A,$

d) $
   (\varphi,\Pi^A\chi)=(\Pi^{A+}\varphi,\chi).
   (\varphi,\Lambda^A\chi)=-(\Lambda^{A+}\varphi,\chi).
    $

Elements of $\cal H$ are  called half-densities.
}

To any element of $\varphi\in{\cal D}$ one can assign the following element
of ${\cal L}_F$:
\begin{equation}
\rho_{\varphi}(A)=(\varphi,\Pi^A \varphi).
\end{equation}
It is important that the norm
$||\rho_{\varphi_1}-\rho_{\varphi_2}||$
is small when
 $||\varphi_1-\varphi_2||$
  is small.

{\bf Lemma 18.}
{\it
The following relation is satisfied:
\begin{equation}
||\rho_{\varphi_1}-\rho_{\varphi_2}|| \le
||\varphi_1-\varphi_2||
(||\varphi_1||+||\varphi_2||)
\end{equation}
}

{\bf Proof}. One has
\begin{equation}
||\rho_{\varphi_1}-\rho_{\varphi_2}|| =
\sup_{p(A)=1}|(\varphi_1,\Pi^A\varphi_1)-(\varphi_2,\Pi^A\varphi_2)|
\end{equation}
Notice that $\rho_{\varphi}\in {\cal L}_+$ when $||\varphi||=1$. therefore,
$|\rho_{\varphi}(\Pi^A)|=|(\varphi,\Pi^A\varphi)|\le 1$ in this case.
Thus, $||\pi^A||\le 1$ when $p(A)=1$. Eq. (113) implies then eq.(112).
Lemma 18 is proved.

Let $\rho \in {\cal L}_F$. Denote by $\{ H,\rho\}$ the following functional
$$
\{ H,\rho\}(A)=\rho(\{A,H\}).
$$

{\bf Definition 5.}
{\it
Let $H \in {\cal A},H=H^+$. An abstract Liouville equation is the following
equation for $\rho^t \in {\cal L}_F, t\in {\bf R}$:
\begin{equation}
\frac{d\rho^t}{dt}=\{ H, \rho^t \}.
\end{equation}
An abstract half-density equation for $\varphi^t \in {\cal H}$ is
\begin{equation}
\frac{d\varphi^t}{dt}=\{ H, \varphi^t \}.
\end{equation}
}

{\bf Lemma 19.}
{\it
Let $\varphi^t$ obey eq.(115). Then $\rho_{\varphi^t}$ obey eq.(114).
}

{\bf Proof.} Let $A \in {\cal A}$. One has
$$
\frac{d}{dt} \rho_{\varphi^t}(A)=(\varphi^t,\Pi^A\Lambda^H\varphi^t)+
(\Lambda^H\varphi^t,\Pi^A\varphi^t)=
$$
$$
=(\varphi^t,[\Pi^A,\Lambda^H]\varphi^t)=(\varphi^t, \Pi^{\{A,H\}} \varphi^t)
=\rho_{\varphi^t} (\{A,H\}).
$$
 Therefore, eq.(114) is satisfied.
Lemma 19 is proved.

Let us give examples of half-density representations.

{\bf Example 1.} Consider the Hamiltonian algebra ${\cal A}^c_n$.
Define  oprations $AB\equiv \Pi^A B,\{A,B\}\equiv \Lambda^A B$, where
$A\in {\cal A}_n^c,B\in L^2({\bf R}^{2\nu n})$ by eq. (109). Let ${\cal D}=
{\cal S}({\bf R}^{2\nu n)}$. The axioms of the half-density representation
are satisfied. We can also notice that the element
 $\rho_{\varphi}\in {\cal L}_F$ is equal to
 $$
 \rho_{\varphi}(X)=|\varphi(X)|^2.
 $$
for example, if $\varphi$ is real, $\rho$ is presented as a square of the
function $\varphi$. Because of this reason, the function $\varphi$ has been
called half-density function in ref.[11].

{\bf Example 2.}
Consider the Hamiltonian algebra ${\cal A}_n^q$. Let ${\cal H}$ be
Hilbert space $L^2({\cal X}^n)$. Determine operators
$\Pi^A,\Lambda^A$ as
\begin{equation}
\Pi^A \varphi=A\varphi, L^A \varphi=\frac{1}{i\hbar} A\varphi.
\end{equation}
It is not hard to check axioms H1-H3.The half-densities play the role of the
wave functions in this example. The density matrix $\rho_{\varphi}$ is
proportional to the projection operator on one-dimensional subspace.

{\bf Example 3.} Consider the same Hamiltonian algebra ${\cal A}_n^q$.
Choose the Hilbert space $\cal H$ as the space of Hilbert-Schmidt operators
on $L^2({\cal X}^n)$ with the inner product
$(\varphi,\chi)=Tr \varphi^+\chi.$ Let $\Pi^A,\Lambda^A$ be the following:
\begin{equation}
\Pi^A \varphi=A\varphi, L^A \varphi=\frac{1}{i\hbar} [A,\varphi].
\end{equation}
The density matrix $\rho_{\varphi}$ is $\rho_{\varphi}=\varphi\varphi^+$;
the matrix $\varphi$ has been called half-density [13] for reasons
analogous to example 1.

We can notice that different equations (Schr\"{o}dinger and Wigner)
 are treated from the same point of view: they are abstract half-density
equations for the cases of examples 2 and 3 correspondingly. Contrary
to the half-density representations, the Hamiltonian algebras are identical
for examples 2 and 3.

{\bf 4.}
For the simplicity, consider the case when elements $A$ of the Hamiltonian
algebra $\cal A$ are presented as functions $A_Y$ on a measure space
$\cal Y$, while the algebra operations are as follows:
$$
(AB)_X=\int dYdZ d_{XYZ} A_Y B_Z,
\{ A,B\}_X=\int dYdZ f_{XYZ} A_Y B_Z,
$$
\begin{equation}
\end{equation}
$$
(A^+)_X=A_X^*,X,Y,Z \in {\cal Y}, A,B \in {\cal A},
$$
$f$ and $d$ are (generalized) functions ${\cal Y}\times {\cal Y}
\times {\cal Y} \rightarrow {\bf C}$. To simplify the notations, denote the
integrals like (118) as $d_{XYZ}A_YB_Z$,$f_{XYZ}A_YB_Z$, i.e. we
integrate over repeated indices.

Elements of ${\cal L}_F$ are presented as (generalized) functions on $\cal Y$,
the quantity $\rho(A)$ can be formally written as $\rho(A)=\rho_XA_X$.

A tensor product ${\cal A}^{\otimes N} ={\cal A} \times ...\times
{\cal A}$ is then presented as an algebra of functions
$A_{Y_1...Y_N} : {\cal Y} \times ...\times
 {\cal Y}={\cal Y}^N\rightarrow {\bf C}$, $Y_1,...,Y_N \in {\cal Y}$.
 The algebra operations are defined as:
 $$
 \{A,B\}_{X_1...X_N}= \sum_{p=1}^N d_{X_1Y_1Z_1}...d_{X_{p-1}Y_{p-1}Z_{p-1}}
 f_{X_pY_pZ_p}
$$
$$
\times d_{X_{p+1}Z_{p+1}Y_{p+1}}...d_{X_NZ_NY_N} A_{Y_1...Y_N}
 B_{Z_1...Z_N},
 $$
 $$
 (AB)_{X_1...X_N}= d_{X_1Y_1Z_1}...d_{X_NY_NZ_N} A_{Y_1...Y_N} B_{Z_1...Z_N},
 (A^+)_{X_1...X_N}=A^*_{X_1...X_N},
$$
$$
  A,B\in {\cal A}^{\otimes N},
 X_i,Y_i,Z_i \in {\cal Y}.
 $$
 It is not hard to check axioms of a Hamiltonian algebra for
 ${\cal A}^{\otimes N}$. Note that the axiom A3 is important for such check.

Let $\gamma$ be a polynomial functional $\gamma:{\cal L} \rightarrow {\bf R}$
of the form
\begin{equation}
\gamma[\rho]=\sum_{k=1}^{K_0} \frac{1}{k!} \rho_{X_1}...\rho_{X_k}
W^{(k)}_{X_1...X_k}
\end{equation}
for some observables $W^{(k)}\in {\cal A}^{\otimes k}$.

{\bf Definition 6.}[3]
{\it
An $\epsilon$-uniformization of the functional (119) is a set of observables
$H^{(N)}\in {\cal A}^{\otimes N}:$
\begin{equation}
H^{(N)}_{X_1...X_N}=\sum_{k=1}^{K_0} \frac{\epsilon^{k-1}}{k!}
\sum_{1\le j_1 \ne ...\ne j_k\le N} W^{(k)}_{X_{j_1}...X_{j_k}}
\prod_{p\ne j_i} I_{X_p}.
\end{equation}
}

Consider now tensor products of half-density representations. Without loss
of generality, one can assume that the Hilbert space $\cal H$ is presented
as $L^2({\cal X})$ for some measure space $\cal X$. Elements of $\cal H$
will be denoted as $\varphi_x,x \in {\cal X}$; integrals like
$\int dx \varphi^*_x \chi_x$ will be denoted as $\varphi_x^*\chi_x$.
Let $(L^2({\cal X}),\Pi,\Lambda)$ be a half-density representation of the
Hamiltonian algebra $\cal A$. The operations $\Pi,\Lambda$ are
$$
(A\varphi)_{x}=b_{Xxy}A_X\varphi_y,
\{A,\varphi\}_{x}=c_{Xxy}A_X\varphi_y.
$$
for some (generalized) functions
$b,c:{\cal Y}\times {\cal X}\times {\cal X} \rightarrow {\bf C}$.
Consider the Hilbert space ${\cal H}^{\otimes N}={\cal H}\times...\times
{\cal H} =L^2({\cal X}^N)$ and the following operations:
$$
(A\varphi)_{x_1...x_N}=b_{X_1x_1y_1}...b_{X_Nx_Ny_N}
A_{X_1...X_N}\varphi_{y_1...y_N},
$$
\begin{equation}
\end{equation}
$$
\{A,\varphi\}_{x_1...x_N}=\sum_{p=1}^N b_{X_1x_1y_1}...
b_{X_{p-1}x_{p-1}y_{p-1}}c_{X_px_py_p}
\beta_{X_{p+1}x_{p+1}y_{p+1}}...\beta_{X_Nx_Ny_N}
$$
$$
\times A_{X_1...X_N}\varphi_{y_1...y_N},
$$
where
$
\beta_{Xxy}=b_{Xxy}-i\hbar c_{Xxy}, X_i \in {\cal Y}, x_i,y_i \in \cal X$.
One can check the axioms of definition 4 for the operations (121).
This half-density representation is called a tensor product of $N$
half-density representations $(L^2({\cal X}),\Pi,\Lambda).$

Let $N$ be fixed, $\epsilon=1/N$, $H$ have the form (120). Then the abstract
half-density equaton (115) takes the form
$$
i\frac{d}{dt} \varphi^{N,t}_{x_1...x_N} = \sum_{k=1}^{K_0} \frac{1}{N^{k-1}}
\sum_{1\le l_1<...<l_k \le N} \sum_{q=1}^k W^{(k)}_{S_1...S_k}
b_{S_1x_{l_1}y_{l_1}}...
b_{S_{p-1}x_{l_{p-1}}y_{l_{p-1}}}
ic_{S_qx_{l_q}y_{l_q}}
$$
\begin{equation}
\times\beta_{S_{p+1}x_{l_{p+1}}y_{l_{p+1}}}
..\beta_{S_{k}x_{l_{k}}y_{l_{k}}}
\varphi^{N,t}_{x_1...y_{l_1}...y_{l_k}...x_N}.
\end{equation}
Lemma 19 tells us that the element of ${\cal L}_F$ of the form
$$
\rho^{N,t}_{X_1...X_N}=
(\rho_{\varphi^t})_{X_1...X_N}\equiv b_{X_1x_1y_1}...b_{X_Nx_Ny_N}
\varphi^{t*}_{x_1...x_N}\varphi^t_{y_1...y_N}
$$
obeys the abstract Liouville equation (114).

{\bf Remarks.}

1. The algebras ${\cal A}^c_N$ and ${\cal A}^q_N$ considered above are the
following tensor powers,
${\cal A}_N^c=({\cal A}_1^c)^{\otimes N},
{\cal A}_N^c=({\cal A}_1^c)^{\otimes N}.$
The $N$-th tensor powers of half-density representations (116) and (117) of
the algebra ${\cal A}_1^q$ are half-density representations of the
algebra ${\cal A}^q_N$ which are given by the same equations, (116), (117).
The half-density representation of the algebra ${\cal A}_N^c$ which has been
considered in example 1 is also the $N$-th tensor power of the analogous
representation of algebra ${\cal A}_1^c$.

2. Eq.(114) has the form of the Liouville equation for the case of
algebra ${\cal A}_N^c$ and of the Wigner equation for algebra ${\cal A}_N^q$.
For the case of the system of $N$ particles moving in the external potential
$U$ and interacting each other with the potential $\frac{1}{N}V$, the
observable $H$ entering to eq.(114) has the form of uniformization (120) of
some functional $\gamma$ which has the same form
$$
\gamma[\rho]=\int dpdq \rho(p,q) \left( \frac{p^2}{2m} +U(q)\right)
$$
$$
+ \frac{1}{2} \int dp_1dp_2 dq_1dq_2 V(q_1,q_2) \rho(p_1,q_1)\rho(p_2,q_2)
$$
both for the classical and quantum cases.
One should only take into account in quantum case that $\rho(p,q)$
is a symbol of the operator $\rho\in {\cal L}$ in $L^2({\bf R}^{\nu})$
which is defined as
$$
\rho(p,q)=\frac{1}{(2\pi \hbar)^{\nu}}\int dy \rho_K(x,y)e^{\frac{i}{\hbar}
p(y-x)},
$$
where $\rho_K$ is a kernel of the operator $\rho$.

3.For the case of the Hamiltonian (120), eq.(114) has been
considered in [3]. It was shown that the property of the correlation
functions
\begin{equation}
{\cal R}^{(N,k,t)}_{X_1...X_k}=\rho^{N,t}_{X_1...X_N}I_{X_{k+1}}...I_{X_N}
\end{equation}
to be approximately equal to the products of one-particle correlators
\begin{equation}
{\cal R}^{(N,k,t)}_{X_1...X_k}
\rightarrow_{N\rightarrow\infty} \rho^t_{X_1}...\rho^t_{X_k}
\end{equation}
conserves under time evolution, while $\rho_X$ obeys the abstract Vlasov
equation
\begin{equation}
\frac{d\rho_X^t}{dt}=f_{YXZ}
 \frac{\partial \gamma}{\partial \rho_Z}
(\rho^t)
\rho^t_Y.
\end{equation}
This justifies the chaos conservation hypothesis.

{\bf 5.} One can notice that eq.(122) for the half-density is of the type
(64), if $H_N=NH^N_0$,$H_0^N$ has the form (63) and
$$
H_0(\varphi^*,\varphi)= \sum_{k=1}^{K_0} \frac{1}{k!}
\sum_{q=1}^k W^{(k)}_{S_1...S_k}
b_{S_1x_{1}y_{1}}...
b_{S_{q-1}x_{q-1}y_{q-1}}
ic_{S_qx_{q}y_{q}}
\beta_{S_{q+1}x_{q+1}y_{q+1}}
..\beta_{S_{N}x_{N}y_{N}}
$$
\begin{equation}
\times\varphi^*_{x_1}...\varphi^*_{x_k}\varphi_{y_1}...\varphi_{y_k}.
\end{equation}
Therefore, the technique developed in sections 5-7 can be applied to eq.(122).
One can construct such function $\varphi^{N,t}_{as} \in L^2({\cal X}^N)$ that
\begin{equation}
(\varphi^{N,t}_{as}-\varphi^{N,t},\varphi^{N,t}_{as}-\varphi^{N,t})
\rightarrow_{N\rightarrow\infty} 0,
\end{equation}
where $\varphi^{N,t}$ is a solution to the Cauchy problem for eq.(122).
Lemma 18 tells us that estimation (127) means that one can use approximate
half-densities in order to find limits as $N\rightarrow\infty$ of average
values of the observables uniformly bounded with respect to $N$. In
particular, one can confirm the chaos conservation hypothesis for
the correlation functions of finite orders and deny it for the $N$-particle
densities in a way analogous to section 3.

It is interesting that eq.(124) is the analog of the Ehrenfest theorem.
Namely, one can introduce creation and annihilation operators $a^{\pm}$
and apply lemma 10. Eq.(120) takes the following form:
\begin{equation}
\frac{i}{N}\frac{d}{dt} \varphi^{N,t} = H_0(a^+/\sqrt{N},a^-/\sqrt{N})
\varphi^{N,t},
\end{equation}
where $\varphi^t$ is such element of $\cal F$ that only $N$-th component
of it differs from zero, while
$$
H_0(a^+/\sqrt{N},a^-/\sqrt{N})= \sum_{k=1}^{K_0} \frac{1}{N^kk!}
\sum_{q=1}^k W^{(k)}_{S_1...S_k}
$$
\begin{equation}
\times b_{S_1x_{1}y_{1}}...
b_{S_{q-1}x_{q-1}y_{q-1}}
ic_{S_qx_{q}y_{q}}
\beta_{S_{q+1}x_{q+1}y_{q+1}}
..\beta_{S_{N}x_{N}y_{N}}
a^+_{x_1}...a^+_{x_k}a^-_{y_1}...a^-_{y_k}.
\end{equation}
The correlation functions
 ${\cal R}^{(N,k,t)}_{X_1...X_k}$
 can be
presented through the average values of the function of operators
$a^{\pm}/\sqrt{N}$:
$$
{\cal R}^{(N,k,t)}_{X_1...X_k}=\frac{(N-k)!}{N!}
(\varphi^{N,t}, b_{X_1x_1y_1}...b_{X_kx_ky_k}a^+_{x_1}...a^+_{x_k}
a^-_{y_1}...a^-_{y_k} \varphi^{N,t}).
$$
As the commutator between operators $a^{\pm}/\sqrt{N}$ tends to zero as
$N\rightarrow \infty$, one obtains eq.(125) from Heisenberg equations.
Thus, one confirms the chaos conservation for correlation functions.

Let us present the theorem for the $N$-particle density
which implies the results of [11,13] for multiparticle
Liouville and Wigner equations. Let $H$ have the
form (120), $\epsilon=1/N$. Consider the solution to eq.(114) that
satisfies the initial condition
$$
\rho^{N,0}_{X_1...X_N}=b_{X_1x_1y_1}...b_{X_Nx_Ny_N}
(K^N_{\varphi^0}\Phi_{R^0})^*_{x_1...x_N}
(K^N_{\varphi^0}\Phi_{R^0})_{y_1...y_N}
$$
Let $\varphi^t$ be a solution to the Cauchy problem for eq.(66), $R^t$
be such a solution to eq.(67) that obeys the initial condition $R^0$.
Denote
$$
c^t=\exp\left(
-\frac{i}{2}\int^t_0 d\tau \frac{\delta^2H_0}{\delta\varphi_x\delta\varphi_y}
R^{\tau}_{xy}\right),
$$
$$
\rho^{N,t,as}_{X_1...X_N}=|c^t|^2 b_{X_1x_1y_1}...b_{X_Nx_Ny_N}
(K^N_{\varphi^t}\Phi_{R^t})^*_{x_1...x_N}
(K^N_{\varphi^t}\Phi_{R^t})_{y_1...y_N}
$$

{\bf Theorem 6.}
{\it
The following relation takes place:
$$
||\rho^{N,t,as}_N -\rho^{N,t}_N || \rightarrow_{N\rightarrow\infty} 0.
$$
}
This theorem is a corollary of theorem 3, lemma 18 and reality of $S^t$
(eq.(38)).

Consider now eq.(66) for our case in more details. Note that it can be treated
as a Hamiltonian system if one introduces the following Poisson
brackets:
$$
\{\varphi_x,\varphi_y\}=\{\varphi_x^*,\varphi_y^*\}=0,
 \{\varphi_x,\varphi_y^*\}=-i.
$$
One can present the functional $H_0$ as
$$
H_0=\frac{1}{\hbar} [\gamma[\rho] -\gamma[\rho-\hbar \sigma]], \hbar \ne 0;
$$
\begin{equation}
\end{equation}
$$
H_0=\frac{\partial \gamma}{\partial \rho_X} \sigma_x , \hbar=0;
$$
where
\begin{equation}
\sigma_X=ic_{Xxy}\varphi^*_x\varphi^*_y, \rho_X=b_{Xxy}\varphi^*_x\varphi^*_y.
\end{equation}
The Poisson brackets between $\rho,\sigma$ are the following:
$$
\{\rho_X,\rho_Y\}=\hbar f_{ZXY}\rho_Z,
\{\rho_X,\sigma_Y\}= f_{ZXY}\rho_Z,
\{\sigma_X,\sigma_Y\}= f_{ZXY}\sigma_Z.
$$
Therefore, the Hamiltonian system in terms of $\rho,\sigma$ is divided
in quantum case into two independent parts:
as $\{\rho_X,\rho_Y-\hbar\sigma_Y\}=0$, the equations for
$\rho$ and $\rho-\hbar\sigma$ are independent. In classical case
the Hamiltnian system consists of two equations: the equation for
$\rho$ coincides with eq.(122), another equation for $\sigma$ is linear.

\section{Conclusions}

We have developed a new asymptotic method that allows us to find
approximations for functions of a large number $N$ of arguments as
$N\rightarrow\infty,$ as well as the corrections to the leading order
of the asymptotic formula. We have seen that this technique is applicable to
eq.(64) being of a general form, as well as to the set (finite or infinite)
of such equations. Multiparticle Schr\"{o}dinger, Liouville
and Wigner equations are partial cases of eq.(64).

We have noticed that for the case of Schr\"{o}dinger equation our approximate
wave function can be used instead of the exact wave function for finding
limits of mean values of general observables uniformly bounded with respect
to $N$. We have considered the case of a general Hamiltonian algebra,
justified the chaos conservation hypothesis for the correlators and
denied it for the $N$-particle densities. It is interesting that for the
operator-valued case the chaos does not conserve even for the correlators.

Two methods have been used for constructing such asymptotics. One of them
is heuristic and allows us to construct the asymptotic formula up to a
multiplicative factor $c^te^{iNC^t}$, where constant $C^t$ does not
depend on the solution to the Hartree-like equaion (66), while $c^t$
depend, in general, on this solution. We can see from theorem 3 that the
reason for this arbitrariness is as follows. When we change the operator
$H_1^N$, the Hartree-like equation (66) does not change, while the number
$c^t$ is multiplied by a $\varphi^t$-dependent factor. The function $C^t$
will change when one adds the constant $h^t$ to the operator $H^N_0$.
As the heuristic method of section 4 uses only the form of
eq.(64), not  the form of the hamiltonian, it cannot predict $c^t,C^t$.

Note that the argumentation of section 4 is also applicable to the
operator-valued case. This implies that the only  difference of the
asymptotic formulas for the ordinary and operator-valued cases is the value of
the constant $c^t$. This conclusion is justified: all quantities entering
to eq.(101) are identical to the analogous quantities in the asymptotic
formula of section 5, except for the phase factor $c^t$.

The relation between the considered technique and complex germ method [10]
can be investigated in more details [6,15-17].  As we have seen in
section 10, one can consider such
representation for the $N$-particle wave function that eq.(64) will
transform into ordinary Schr\"{o}dinger-like equation, while the analog
of the Planck constant will be $1/N$.
The asymptotics constructed in this paper correspond [6,17] to the
isotropic manifold [10] of the special form. General Lagrangian manifolds
with complex germs can be also considered when one studies the infinite
superposition of the obtained asymptotic formulas, see [6,15-17] for
more details.

\section*{Acknowledgements}
This work was supported by RFFI, grant \# 93-012-1075.

\section*{Appendix A}

As mentioned in conclusions, our asymptotic method for finding approximate
 solutions to eq.(64) as $N\rightarrow\infty$ is analogous to the
 complex germ technique. Namely, eq.(64) is analogous to the ordinary
 Schr\"{o}dinger-like equation
 \begin{equation}
 i\hbar \frac{\partial \psi^t(x)}{\partial t}
 = H\left(x,-i\hbar\frac{\partial}{\partial x}\right)\psi^t(x),
  x\in {\bf R}^n,
 \end{equation}
the Hartree-like equation (66) for complex function $\varphi$ is an
analog of the Hamiltonian system
 \begin{equation}
 \frac{dQ^t}{dt}=\frac{\partial H}{\partial P}(Q^t,P^t),
 \frac{dP^t}{dt}=-\frac{\partial H}{\partial Q}(Q^t,P^t),
 \end{equation}
for two real vectors $P^t,Q^t \in {\bf R}^n$.
According to section 10, the chaos conservation
hypothesis for correlation functions resembles the Ehrenfest theorem.
The multiparticle canonical operator is analogous to the canonical operator
for the complex germ in a point, the constructed asymptotic solutions
resemble the wave-packet-like approximate solutions in quantum mechanics
 \begin{equation}
 \psi^t(x)=c^te^{\frac{i}{\hbar}(S^t+P^t(x-Q^t))}
 g^t\left(\frac{x-Q^t}{\sqrt{\hbar}}\right)+O(\sqrt{\hbar}),
 \end{equation}
$g^t \in {\cal S}({\bf R}^n).$

When initial conditions are given, the functions $c^t,S^t,g^t$ can be found
by solving the equations obtained in complex germ theory by subsitution of
eq.(134) to eq.(132).
 The purpose of this appendix is to show how such equations
can be heuristically derived without such substitution in a way analogous
to section 4.

First of all, find the dependence of $S^t$ on $P^t,Q^t$. When one varies
the initial condition for the classical trajectory
by the quantity of order $\hbar$, vectors $P^t,Q^t$ varies as
$$
P^t \rightarrow P^t+\hbar \delta P^t,
Q^t \rightarrow Q^t+\hbar \delta Q^t,
$$
the only change of the wave function (134) as $\hbar\rightarrow 0$ is
multiplication by the quantity
$$
e^{i(\delta S^t - P^t \delta Q^t)},
$$
where $\delta S^t$ is a variation of the quantity $S^t(P^0,Q^0)$ depending
on the initial condition for the classical trajectory. Therefore,
$$
\delta S^t - P^t \delta Q^t=const,
$$
i.e.
 \begin{equation}
S^t=\int_0^t [P^{\tau}\frac{d}{d\tau} Q^{\tau}-H(P^{\tau},Q^{\tau})]d\tau
+C^t
 \end{equation}
for some quantity $C^t$ that does not depend on the classical trajectory.

Consider the variation of the solution to eq.(133) by a quantity of order
$\sqrt{\hbar}$:
$$
P^t \rightarrow P^t+\sqrt{\hbar} \delta P^t,
Q^t \rightarrow Q^t+\sqrt{\hbar} \delta Q^t,
$$
The function (134) transforms then into the following function:
 \begin{equation}
c^te^{\frac{i}{\hbar}(S^t+P^t(x-Q^t))}
 g^{t'}\left(\frac{x-Q^t}{\sqrt{\hbar}}\right)+O(\sqrt{\hbar}),
 \end{equation}
where
$$
g^{t'}(\xi)=const \exp\left(i
\left(\delta P^t\xi-
\delta Q^t\frac{1}{i}\frac{\partial}{\partial \xi}
\right)\right) g^t(\xi).
$$
As the functions (134),(136) should be asymptotic solutions to eq.(132),
the operator
 \begin{equation}
\delta P^t\xi-
\delta Q^t\frac{1}{i}\frac{\partial}{\partial \xi}
 \end{equation}
should transform solutions to the equation for $g^t$ to solutions, when
$(\delta P^t,\delta Q^t)$ is a solution to the variation system
$$
\frac{d}{dt}\delta Q^t =\frac{\partial^2H}{\partial P\partial P} \delta P^t+
\frac{\partial^2H}{\partial P\partial Q} \delta Q^t,
$$
 \begin{equation}
 \end{equation}
$$
-\frac{d}{dt}\delta P^t =\frac{\partial^2H}{\partial Q\partial P} \delta P^t+
\frac{\partial^2H}{\partial Q\partial Q} \delta Q^t,
$$
Notice that $(\delta P^t,\delta Q^t)$ may be complex quantities.

Therefore, if
$
(\delta P^0\xi-
\delta Q^0\frac{1}{i}\frac{\partial}{\partial \xi})g^0=0
$
then
 \begin{equation}
(\delta P^t\xi-
\delta Q^t\frac{1}{i}\frac{\partial}{\partial \xi})g^t=0.
 \end{equation}
The conservation of property (139) under time evolution allows us to
introduce a notion of complex germ.

Let $\alpha_{ij}$ be symmetric complex matrix $n\times n$ such that
$Im \alpha >0$. Consider the following function $g_{\alpha}\in
{\cal S}({\bf R}^n)$:
 \begin{equation}
 g_{\alpha}(\xi)=\exp\left(\frac{i}{2}\sum_{k,l=1}^n \xi_k\alpha_{kl}\xi_l
 \right)
 \end{equation}
The following definition is analogous to definition 2.

{\bf Definition 7} [10].
{\it The following $n$-dimensional subspace of the complex $2n$-dimensional
space ${\bf R}^n$:}
$$
{\cal G}_{\alpha}
=\{ (p_1,...,p_n;q_1,...,q_n)| p_i=\sum_{j=1}^n \alpha_{ij}q_j \}
$$
{\it will be referred to as a complex germ corresponding to the matrix
$\alpha$.}

The following lemma is the analog of lemma 8.

{\bf Lemma 20.}
{\it
1. Let $(p,q) \in {\cal G}_{\alpha}.$ Then
$$
(p\xi-
q\frac{1}{i}\frac{\partial}{\partial \xi})g_{\alpha}=0
$$
 for arbitrary $(p,q)\in {\cal G}_{\alpha}$.

 2.Let
 $
(p\xi-
q\frac{1}{i}\frac{\partial}{\partial \xi})f=0.
$
Then $f=cg_{\alpha}$ for some constant $c\in {\bf C}$.
}

The proof is straightforward.

Let $(\delta P^0,\delta Q^0)\in {\cal G}_{\alpha^0}.$ Then
 $(\delta P^t,\delta Q^t)\in {\cal G}_{\alpha^t}$
 for the matrix $\alpha^t$ being a solution to the Riccati equation:
 \begin{equation}
\frac{d}{dt}\alpha^t=
-\frac{\partial^2 H}{\partial Q\partial Q}
-\frac{\partial^2 H}{\partial Q\partial P}\alpha^t
-\alpha^t\frac{\partial^2 H}{\partial P\partial Q}
-\alpha^t\frac{\partial^2 H}{\partial Q\partial Q}\alpha^t.
 \end{equation}
Lemma 20 and eq.(139) tell us that the initial condition $g^0=g_{\alpha^0}$
evolve into the following function
$$
g^t=c^tg_{\alpha^t}.
$$
for some constant $c^t\in {\bf C}$. Thus, we see that Gaussian wave packet
(134) evolve into a Gaussian one as $\hbar \rightarrow 0.$ One can also
consider complex germ creation operators of the form (139), where
$(\delta P^{t*},\delta Q^{t*})\in {\cal G}_{\alpha^t}$ and find another
asymptotic solutions to eq.(132). Thus, the presented approach allows us to
find unambiguously complex germ creation and annihilation operators and
the Riccati equation (141) without consideration of eq.(132).

\section*{Appendix B}

In this appendix we show how one can reproduce for the operator-valued case
the classical Hamiltonian system (133), as well as the Hartree-like equation
(66) by making use of the equations for mean values of the observables and of
the argumentation analogous to the derivation of the Ehrenfest theorem
(see, for example, [20]). Note also that the method to be presented in this
appendix is heuristic.

{\bf 1.} Consider the ordinary quantum mechanical Schr\"{o}dinger equation
\begin{equation}
i\hbar \frac{\partial \psi^t(Q,q)}{\partial t} =
H \left(Q,-i\hbar\frac{\partial}{\partial Q},
q,-i\frac{\partial}{\partial q}\right)
\psi^t(Q,q)
\end{equation}
corresponding to the physical system which is ''semiclassical'' with respect
to $Q\in {\bf R}^n$ and ''quantum'' with respect to $q\in {\bf R}^m$. Consider
such solutions $\psi^t \in L^2({\bf R}^{n+m})$ that
\begin{equation}
\left(\psi^t,
A\left(Q,-i\hbar\frac{\partial}{\partial Q},
q,-i\frac{\partial}{\partial q}\right)
\psi^t\right)-
\left(\psi^t,
A\left(Q^t,P^t,q,-i\frac{\partial}{\partial q}\right)
\psi^t \right)
\rightarrow_{\hbar\rightarrow 0} 0
\end{equation}
for $\hbar$-independent functions $A$.
For example, the mean values of ''semiclassical'' observables
$A(Q,-i\hbar\frac{\partial}{\partial Q})$
are required to have limits as $\hbar \rightarrow 0$
$$
\left(\psi^t,A\left(Q,-i\hbar\frac{\partial}{\partial Q}\right)
\psi^t\right)
\rightarrow A(Q^t,P^t).
$$
Notice that there exist wave functions obeying the condition (143):
examples of them are functions like
$$
\psi^t=e^{\frac{i}{\hbar}(S^t+P^t(Q-Q^t))}
g^t\left(q,\frac{Q-Q^t}{\sqrt{\hbar}}\right),
$$
where $g^t\in {\cal S}({\bf R}^{n+m})$.

We are to find possible equations for $P^t,Q^t$. To do this, consider first
the equation for the mean value of some observable
$\hat{B}=B(q,-i\frac{\partial}{\partial q})$:
\begin{equation}
i\hbar \frac{\partial}{\partial t}
(\psi^t,\hat{B}\psi^t)=
(\psi^t,[\hat{B},H]\psi^t),
\end{equation}
where
$H=
H(Q,-i\hbar\frac{\partial}{\partial Q},q,-i\frac{\partial}{\partial q})$.
As the left-hand side of eq.(144) is of order $O(\hbar)$ as $\hbar
\rightarrow 0$, the right-hand side of it should also vanish in a
leading order of $\hbar$. Eq.(143) implies that one should replace
$Q$ by $Q^t$ and $-i\hbar\partial/\partial Q$ by $P^t$ for calculating
the leading order of the right-hand side of eq.(144).

Therefore, one should demand
\begin{equation}
(\psi^t,[\hat{B},\hat{H}]\psi^t)=0,
\end{equation}
where
$\hat{H}=
H(Q^t,P^t,q,-i\frac{\partial}{\partial q})$.
Eq.(145) implies that for arbitrary $\hat{B}$
$Tr(\hat{B}[\hat{H},\Pi^t])=0,$
where $\Pi^t$ is a projector on $\psi^t$. This implies that
$[\hat{H},\Pi^t]=0$,i.e.
$$
\hat{H}\psi^t=\lambda\psi^t
$$
for some $\lambda=\lambda(P^t,Q^t)\in{\bf R}.$

Consider now the equation for the mean value of the observable
$A(Q,-i\hbar\frac{\partial}{\partial Q})$.
Making use of eq.(143), one finds
$$
i\hbar\frac{dA}{dt}=
\left(\psi^t, i\hbar
\left(\frac{\partial A}{\partial Q^t}\frac{\partial \hat{H}}{\partial P^t}-
\frac{\partial A}{\partial P^t}\frac{\partial \hat{H}}{\partial Q^t}
\right)
\psi^t\right)+O(\hbar^2),
$$
where $A=A(Q^t,P^t)$. We have taken into account that the semiclassical
approximation for the commutator of two ''semiclassical'' observables
is [21] their Poisson bracket multiplied by $i\hbar$. By using the
relation
\begin{equation}
(\psi^t,\delta \hat{H}\psi^t)=\delta\lambda
\end{equation}
one has
$$
\frac{dA}{dt}=
\frac{\partial A}{\partial Q^t}\frac{\partial \lambda}{\partial P^t}-
\frac{\partial A}{\partial P^t}\frac{\partial \lambda}{\partial Q^t}.
$$
Therefore, we have reproduced the Hamiltonian system being used in
constructing asymptotics for the operator-valued case [14].

{\bf 2.} An analogous technique can be also used for derivation of eq.(66)
for the operator-valued case. Let us illustrate the approach for the case
of constructing asymptotics for eq.(107).

First of all, let us represent eq.(107) through the creation and annihilation
operators. Analogously to section 6, consider such Hilbert space ${\cal H}$
that
${\bf R}^2\otimes L^2({\bf R}^{\nu n})\subset {\cal H}$.
We choose $\cal H$ as a space of sets of functions
$$
\Phi_{n,m}:{\bf R}^{\nu}\times ... \times {\bf R}^{\nu} \times
\{1,2\} \times ... \times \{1,2\} \equiv {\bf R}^{\nu n} \times \{1,2\}^m
\rightarrow {\bf C}
$$
which satisfy the condition
$$
\sum_{n,m=0}^{\infty} \sum_{I_1,...,I_m=1}^2 \int dx_1...dx_n
|\Phi_{n,m}(x_1,...,x_n,I_1,...,I_m)|^2 < \infty
$$
and are symmetric separately with respect to $x_i$ and with respect to
$I_i$. We identify such element $\Phi \in {\cal H}$ that

a) $\Phi_{n,m}=0$ as $n\ne N$ or $m \ne 1$;

b) $\Phi_{n,1}(x_1,...,x_N,I)=\Psi_{N,I}(x_1,...,x_N)$

with the element
$\Psi \in {\bf R}^2\otimes L^2({\bf R}^{\nu n})$.
By $\hat{\Pi}$ we denote the projector on the subspace
${\bf R}^2\otimes L^2({\bf R}^{\nu n})$ of the Hilbert space
$\cal H$.

We introduce the following creation and annihilation operators in $\cal H$:
$$
(a^+(x)\Phi)_{n,m}(x_1,...,x_n,I_1,...,I_m)=
$$
$$
=\frac{1}{\sqrt{n}}
\sum_{i=1}^n \delta(x-x_i)
\Phi_{n-1,m}(x_1,...,x_{i-1},x_{i+1},...,x_n,I_1,...,I_m),
$$
$$
(a^-(x)\Phi)_{n-1,m}(x_1,...,x_{n-1},I_1,...,I_m)=\sqrt{n}
\Phi_{n,m}(x,x_1,...,x_{n-1},I_1,...,I_m),
$$
$$
(b^+_I\Phi)_{n,m}(x_1,...,x_n,I_1,...,I_m)=\frac{1}{\sqrt{m}}
\sum_{j=1}^m \delta_{I I_j}
\Phi_{n,m-1}(x_1,...,x_n,I_1,...,I_{j-1},I_{j+1},...,I_m),
$$
$$
(b^-_I\Phi)_{n,m-1}(x_1,...,x_{n},I_1,...,I_{m-1})=\sqrt{m}
\Phi_{n,m}(x,x_1,...,x_{n},I,I_1,...,I_{m-1}),
$$
where $x\in{\bf R}^{\nu},I\in\{1,2\}$.

Consider the operator $H:{\cal H}\rightarrow {\cal H}$ of the form
$$
NH(a^+/\sqrt{N},a^-/\sqrt{N},b^+,b^-)=\int dx a^+(x)a^-(x)
\sum_{I,J=1}^2 b_I^+ B_{IJ}(x)b^-_J +
$$
$$
\frac{1}{\hbar} (\int dx a^+(x) [-\frac{\hbar^2}{2m}\Delta +U(x)]a^-(x)+
$$
$$
+\frac{1}{2N} \int dx dy a^+(x)a^+(y) V(x,y) a^-(x)a^-(y) )
\sum_{I=1}^2 b_I^+ b_I^-.
$$
Analogously to lemma 10, one shows that any solution to eq.(107) obeys also
the following equation
\begin{equation}
(i\frac{\partial}{\partial t}-NH)\Psi^t =0.
\end{equation}
Let us analyse this equation in a way analogous to the consideration of
eq.(142). Consider such solutions to eq.(147) that
$$
(\Psi^t, A(a^+/\sqrt{N},a^-/\sqrt{N},b^+,b^-)\Psi^t) -
$$
\begin{equation}
-(\Psi^t, A(\varphi^{t*},\varphi^t,b^+,b^-)\Psi^t)
\rightarrow_{N\rightarrow\infty} 0
\end{equation}
for any polynomial function $A$ being invariant under substitution
$a^{\pm}(x)\rightarrow a^{\pm}(x)e^{\pm i\alpha}$,
$b^{\pm}_I\rightarrow b^{\pm}_Ie^{\pm i\beta}$, $\alpha,\beta=const.$
Example of $\Psi^t$ obeying eq.(149) is
$$
\Psi^t_{N,I}=K^N_{\varphi^t} g^t \zeta_I.
$$
Let us derive now eq. (108). Consider the mean value of the operator
$\hat{B}=B(b^+,b^-).$ One has:
$$
i\frac{d}{dt} (\Psi^t,\hat{B}\Psi^t)=N(\Psi^t,[\hat{B},H]\Psi^t).
$$
As the right-hand side of this equation is of order $O(N)$ as
$N\rightarrow\infty$, one should require it to vanish. Analogously to
the quantum mechanical case, this implies that
$$
(H(\varphi^{t*},\varphi^t,b^+,b^-)-\lambda)\Psi^t=0.
$$
Making use of the definition of the operator $H$, one shows
that when
$\Psi \in {\bf R}^2\otimes L^2({\bf R}^{\nu n})$, there are two
eigenvalues $\lambda$,
$$
\lambda = H_0^{\pm} (\varphi^{t*},\varphi^t,b^+,b^-),
$$
and
$$
\Psi^t_{N,I,\pm}=\zeta_I^{\pm}(\varphi^{t*},\varphi^t) X_{N,\pm}^t.
$$
These eigenvalues and eigenfunctions coincide with those which has been
found in section 9.

As the commutator between operators $a^{\pm}/\sqrt{N}$ tends to zero
as $N\rightarrow\infty$, one can use ordinary semiclassical technique
[21] to compute the commutator $[A,H]$. Let
$$
A=A(a^+/\sqrt{N},a^-/\sqrt{N}).
$$
Then
$$
(\Psi^t,[A,H]\Psi^t) -
$$
$$
- \left(\Psi^t, \int dx \left(
\frac{\delta A}{\delta \varphi(x)}
\frac{\delta H}{\delta \varphi^*(x)} -
\frac{\delta H}{\delta \varphi(x)}
\frac{\delta A}{\delta \varphi^*(x)}
\right) \Psi^t \right) \rightarrow_{N\rightarrow\infty} 0,
$$
we have omitted the arguments $\varphi^*,\varphi$ of the function $A$
and the arguments $\varphi^*,\varphi,b^+,b^-$ of the function $H$.
Making use of eq.(146), one obtains
\begin{equation}
i\frac{dA}{dt} =\int dx \left(
\frac{\delta A}{\delta \varphi(x)}
\frac{\delta H_0^{\pm}}{\delta \varphi^*(x)} -
\frac{\delta H_0^{\pm}}{\delta \varphi(x)}
\frac{\delta A}{\delta \varphi^*(x)}
\right),
\end{equation}
where the sign $+$ or $-$ depends on the choice of the wave function
$\Psi^t$. We can notice that eq.(149) is a consequence of the Hartree-like
equation (108). Thus, the approach based on the Ehrenfest theorem allows us
to derive the equation for $\varphi^t$.

\end{document}